\documentclass{elsarticle}

\usepackage{float}

\usepackage{booktabs} % To thicken table lines

\usepackage{xcolor}
\usepackage{listings}
\definecolor{mGreen}{rgb}{0,0.6,0}
\definecolor{mGray}{rgb}{0.5,0.5,0.5}
\definecolor{mPurple}{rgb}{0.58,0,0.82}
\definecolor{backgroundColour}{rgb}{0.95,0.95,0.92}

\lstdefinestyle{CStyle}{
    backgroundcolor=\color{backgroundColour}, 
    commentstyle=\color{mGreen},
    keywordstyle=\color{magenta},
    numberstyle=\tiny\color{mGray},
    stringstyle=\color{mPurple},
    basicstyle=\footnotesize,
    breakatwhitespace=false,         
    breaklines=true,                 
    captionpos=b,                    
    keepspaces=true,                 
    numbers=left,                    
    numbersep=5pt,                  
    showspaces=false,                
    showstringspaces=false,
    showtabs=false,                  
    tabsize=2,
    language=C
}

\usepackage{newtxtext}
\usepackage{amsmath}
\DeclareMathAlphabet{\mathcal}{OMS}{cmsy}{m}{n}
\usepackage{bm} % bold math
\usepackage{amssymb} % For number sets symbols

\RequirePackage[left=20mm,right=20mm,top=20mm,bottom=20mm]{geometry}

\usepackage{algorithm}% http://ctan.org/pkg/algorithms
\usepackage{algpseudocode}% http://ctan.org/pkg/algorithmicx
\usepackage{subcaption}
\usepackage{hyperref} %	 hyperlinks

\usepackage{siunitx}

\graphicspath{ {images/} }

\journal{Computer Methods in Applied Mechanics and Engineering}

\usepackage{numcompress}
\begin{document}

% declaration of the 'ParFor'
\algblock{ParFor}{EndParFor}
% customising the new block
\algnewcommand\algorithmicparfor{\textbf{parfor}}
\algnewcommand\algorithmicpardo{\textbf{do}}
\algnewcommand\algorithmicendparfor{\textbf{end\ parfor}}
\algrenewtext{ParFor}[1]{\algorithmicparfor\ #1\ \algorithmicpardo}
\algrenewtext{EndParFor}{\algorithmicendparfor}

% declaration of 'Barrier'
\algblock{Barrier}{EndBarrier}
% customising the new block
\algnewcommand\algorithmicbarrier{\textbf{barrier}}
\algnewcommand\algorithmicendbarrier{\textbf{continue}}
\algrenewtext{Barrier}{\algorithmicbarrier}
\algrenewtext{EndBarrier}{\algorithmicendbarrier}

\begin{frontmatter}

\title{\texttt{PeriPy} - A High Performance \texttt{OpenCL} Peridynamics Package}

\author[mymainaddress,mysecondaryaddress]{B. Boys\corref{mycorrespondingauthor}}
\cortext[mycorrespondingauthor]{Corresponding author}

\author[mysecondaryaddress,mytertiaryaddress]{T. J. Dodwell}

\author[mymainaddress]{M. Hobbs}

\author[mymainaddress,mytertiaryaddress]{M. Girolami}

\address[mymainaddress]{Department of Engineering, University of Cambridge, Trumpington Street, Cambridge CB2 1PZ, UK}
\address[mysecondaryaddress]{College of Engineering, Mathematics and Physical Sciences, University of Exeter, UK}
\address[mytertiaryaddress]{The Alan Turing Institute, London, NW1 2DB, UK}

\begin{abstract}
This paper presents a lightweight, open-source and high-performance python package for solving peridynamics problems in solid mechanics. The development of this solver is motivated by the need for fast analysis tools to achieve the large number of simulations required for {\em `outer-loop'} applications, including  sensitivity analysis, uncertainty quantification and optimisation. Our python software toolbox utilises the heterogeneous nature of \texttt{OpenCL} so that it can be executed on any platform with CPU or GPU cores. We illustrate the package use through a range of industrially motivated examples, which should enable other researchers to build on and extend the solver for use in their own applications. Step improvements in execution speed and functionality over existing techniques are presented. A comparison between this solver and an existing \texttt{OpenCL} implementation in the literature is presented, tested on benchmarks with hundreds of thousands to tens of millions of nodes. We demonstrate the scalability of the solver on the GeForce RTX 2080 TiGPU from NVIDIA, and the memory-bound limitations are analysed. In all test cases, the implementation is between 1.4 and 10.0 times faster than a similar existing GPU implementation in the literature. In particular, this improvement has been achieved by utilising local memory on the GPU.
\end{abstract}

\begin{keyword}
peridynamics \sep non-local continuum \sep fracture \sep Graphics Processing Unit \sep \texttt{OpenCL}
%\MSC[2010] 00-01\sep  99-00
\end{keyword}

\end{frontmatter}

% \linenumbers

\section{Introduction}

Modelling the initiation and propagation of fractures remains a central and classical scientific challenge in computational mechanics. The complex nonlinearities, stress singularities and multiple scales involved with such problems mean that the approaches are extremely computationally intensive for any problem of industrial interest, and there are significant modelling uncertainties. Peridynamic theory, introduced by Silling in 2000 \cite{Silling2000}, provides a promising theoretical framework for developing robust numerical models that can be applied to fracture simulation across a broad range of materials \cite{Javili2019, Diehl2019}. In this contribution, we develop a scalable, high-performance solver exploiting modern \emph{Graphics Processing Unit} (GPU) / \emph{Central Processing Unit} (CPU) architecture, which is readily extended, to ensure that the computational demands of peridynamics simulators do not limit their adoption and future development for real-world problems.

\smallskip
Peridynamics is a non-local theory of continuum mechanics that is well suited to modelling crack initiation and propagation. The peridynamic model defines the material behaviour at a point in a continuum body as an integral equation of the surrounding displacement. This is in contrast to the classical theory of solid mechanics, where material behaviour at a given point is defined by partial differential equations. The classical theory is only valid if the body under analysis has a spatially continuous and twice differentiable displacement field. Spatial derivatives are not defined across discontinuities, and additional techniques are required for fracture modelling. Peridynamic theory does not include spatial derivatives and remains valid across discontinuities. Fracture is an emergent behaviour of the peridynamic governing equation, and no additional assumptions or techniques are required. There are two primary formulations of peridynamic theory: bond-based \cite{Silling2000} and state-based theory \cite{Silling2007}. In the original bond-based theory, pairwise particles interact through a single bond. It is an oversimplification to assume that any pair of particles interact only through a central potential that is independent of all other local conditions. For example, this assumption limits the Poisson ratio to a fixed value. \citet{Silling2007} later introduced a generalised state-based theory that overcomes the limitations of the original theory.

\smallskip
Peridynamics has been successfully applied to a wide range of material failure problems, and \citet{Javili2019} performed a thorough review of peridynamic theory and its many applications for static and dynamic problems across a range of materials. \citet{Diehl2019} recently reviewed benchmark experiments for the validation of peridynamic models. The authors identified 39 publications that compare numerical predictions from peridynamic simulations against experimental data. Validation has been restricted to relatively simplistic element-level benchmark tests, such as the Split Hopkinson Pressure Bar test, Kalthoff--Winkler experiment, three-point flexural tests and anchor bolt pullout. Validation through large-scale experiments is still needed. 

\smallskip
The computationally demanding nature of peridynamics is a major barrier to tackling large-scale industry-motivated problems. Peridynamic models are computationally expensive due to the non-local nature of the governing equations. Simulating large problems of interest, on the order of hundreds of thousands to tens of millions of particles, requires vast computational resources. Reducing the computational expense of peridynamic models is essential for wider adoption and further development. The high computational expense of peridynamic simulations is a major factor preventing detailed validation studies. This is manifested as follows: (1) The vast majority of papers only consider two-dimensional problems. (2) Very few papers address large scale industry-motivated problems. (3) The majority of papers calibrate model parameters against a single experiment, and only a few addressed further validation \cite{Diehl2019}. The availability of a fast solver would enable the development of `outer-loop' applications that require a large number of simulations, including sensitivity analysis, uncertainty quantification \cite{Diehl2016, Franzelin14}, optimisation \cite{Sohouli2020, Kefal2019} and convergence \cite{Bobaru2011} studies. It is essential to quantify the uncertainty in model outputs for peridynamics to become a viable method for addressing industry-motivated problems.

\smallskip
A number of strategies are available to minimise computational effort: (1) employ a multi-scale disretisation scheme and refine the mesh in areas of interest \cite{Ren2016, Hu2018}; (2) adaptively refine the mesh at crack tips \cite{Bobaru2009, Bobaru2011, Dipasquale2014, Gu2017}; (3) couple peridynamic and finite element meshes \cite{Liu2012FEM, Galvanetto2016}; and (4) exploit parallel computing techniques and multi-thread processing. Peridynamics lends itself well to parallel processing techniques because the forces of the bonds and particles can be computed in parallel. Peridynamics is particularly suited to parallelisation when compared to other particle methods, such as Smoothed Particle Hydrodynamics (SPH), because the neighbourhood of nodes remains unchanged throughout the simulation, and so a computationally expensive neighbour search over nodes only needs to be performed once for any given model. The next three paragraphs explore (4), the many existing peridynamics codes that have been developed on distributed, shared and massively parallel GPU memory architectures, respectively.\smallskip

Message Passing Interface (\texttt{MPI}) is the dominant parallel programming model for developing highly scalable codes on distributed memory platforms. \texttt{MPI}-based codes use spatial decomposition, in which the physical space is decomposed into multiple subdomains, and each subdomain is assigned to a unique compute node (each compute node has its own local memory). This approach is used for problems with large memory requirements. A number of peridynamic solvers have been developed that utilise \texttt{MPI} \cite{Silling2003a, Parks2008, Parks2012}. EMU, developed by \citet{Silling2003a}, was the first computational peridynamics code. EMU is a parallel \texttt{MPI}-based Fortran code \cite{Breitenfeld2014}. Peridynamics has been implemented within the classical molecular dynamics package LAMMPS (Large-scale Atomic/Molecular Massively Parallel Simulator). The peridynamics package in LAMMPS is known as PD-LAMMPS \cite{Parks2008}. This was the first open-source peridynamics code. A more recent and capable package is the open-source code Peridigm \cite{Parks2012}. Peridigm is the only available open-source code that is specifically designed for peridynamic simulations. Peridigm is written in C++ and requires many dependencies \cite{PeridigmInstallation}.

\smallskip
The number of cores per compute node has rapidly increased in recent years, and shared memory approaches have been widely adopted. \texttt{OpenMP} (Open Multi-Processing application programming interface) is the standard for developing shared memory multi-threaded code. \texttt{OpenMP} has been used in multiple works \cite{Kilic2008, Fan2017, Lee2017}. Parallelisation with \texttt{OpenMP} is efficient and simple to implement but is constrained by the memory of a single compute node. Hybrid applications can be developed for large-scale clusters, where \texttt{MPI} is used for parallelism across compute nodes and \texttt{OpenMP} is used for parallelism within a multi-core compute node. Hybrid \texttt{MPI}/\texttt{OpenMP} approaches have been employed by \citet{DallaBarba2018} and \citet{Ha2020}. HPX (High Performance ParallelX) has recently been employed by \citet{Diehl2020} to develop scalable peridynamics code.

\smallskip
The use of GPU codes to speed up peridynamics simulations has been limited. Calculation of the bond forces are ideal for GPU acceleration, as each bond calculation can be done on one of the many concurrent threads in a GPU. Mossaiby et al. (2017) \cite{Mossaiby2017} demonstrated that, by replacing the outer loop over \emph{nodes} with a compute kernel executing on each node, a speed increase of a factor of 40 to 90 over optimised sequential C++ code and of 3 to 6 times over \texttt{OpenMP} parallel code can be achieved. The current work improves on the work by Mossaiby et al. \cite{Mossaiby2017} by replacing the outer loop over \emph{bonds} and utilising a binary parallel reduction of the bond forces into the nodal force density by exploiting the execution model and memory model of \texttt{OpenCL}, respectively.

\smallskip
This paper introduces \texttt{PeriPy}, a fast, open-source package with a user interface in Python. The code is hosted on Github \href{https://github.com/alan-turing-institute/PeriPy}{\texttt{https://github.com/alan-turing-institute/PeriPy}} and is fully tested. \texttt{PeriPy} is easy to use, and the reader can get started with the latest documentation at \href{https://peripy.readthedocs.io/en/latest/}{\texttt{https://peripy.readthedocs.io}}. Please see \ref{Installation} for the installation instructions. It is 1.4-10.0x faster than the Mossaiby et al. \texttt{OpenCL} implementation \cite{Mossaiby2017} and therefore opens up the possibility of `Outer-loop' applications including uncertainty quantification, optimisation and feature recognition. \texttt{PeriPy} has support for both regular and irregular mesh files, composite and interface material models, arbitrary n-linear `microelastic' damage models and simulates force or displacement controlled boundary conditions and initial conditions. The user interface allows arbitrary subsets of particles to be easily measured for their state variables. Output files can be viewed in Paraview. Various `partial volume correction' algorithms, `surface correction' algorithms and `micromodulus functions' are included. Various explicit integrators are included, and the code is easily extended to define other higher order and/or adaptive integrators.

\smallskip
It is hoped that this package will become the industry standard and other researchers will contribute to its further development. It is important to have an industry standard solver for a number of reasons: (1) to reduce unnecessary development time; (2) so that research is accessible; (3) so that research is repeatable. Section~\ref{Theory} provides a recap of bond-based peridynamics theory, while Section~\ref{implementation} provides an overview of the code design and the factors that have helped provide the speedup over existing codes. Section~\ref{benchmarking} provides benchmark tests that demonstrate the speedup over existing codes and the code performance on a modern GPU up to tens of millions of nodes, over thousands of time-steps and with varying node family size. Section~\ref{validation} provides a validation of \texttt{PeriPy} against existing codes through an engineering test case, the Kalthoff--Winkler experiment. Section~\ref{concrete_beams} demonstrates the use of the code with a problem from industry, the simulation of notched concrete beam fracture. The experiment involves outer-loop optimisation over peridynamics constitutive model parameters using experimental data from a concrete beam test. The point estimate of the constitutive parameters are tested on experimental data, the first validation of its kind in the literature. Section~\ref{reinforced_concrete_beam} demonstrates the use of the code with another problem from industry, the simulation of a reinforced concrete beam, and shows a visualisation of the cracking pattern. Section~\ref{future} outlines future avenues for developing \texttt{PeriPy}, and Section~\ref{conclusions} concludes the paper.

\section{Preliminaries: Bond-based peridynamics equations and their disretisation}\label{Theory}
In this section, we will briefly review {\em bond-based} peridynamics continuum theory and its numerical disretisation. The implementation presented is readily applicable to state-based theory \cite{Silling2007, Warren2009}, and this connection is briefly discussed in Sec.~\ref{sec:StateBased}. 

\begin{figure}
	\centering    
	\includegraphics[width=14.0cm]{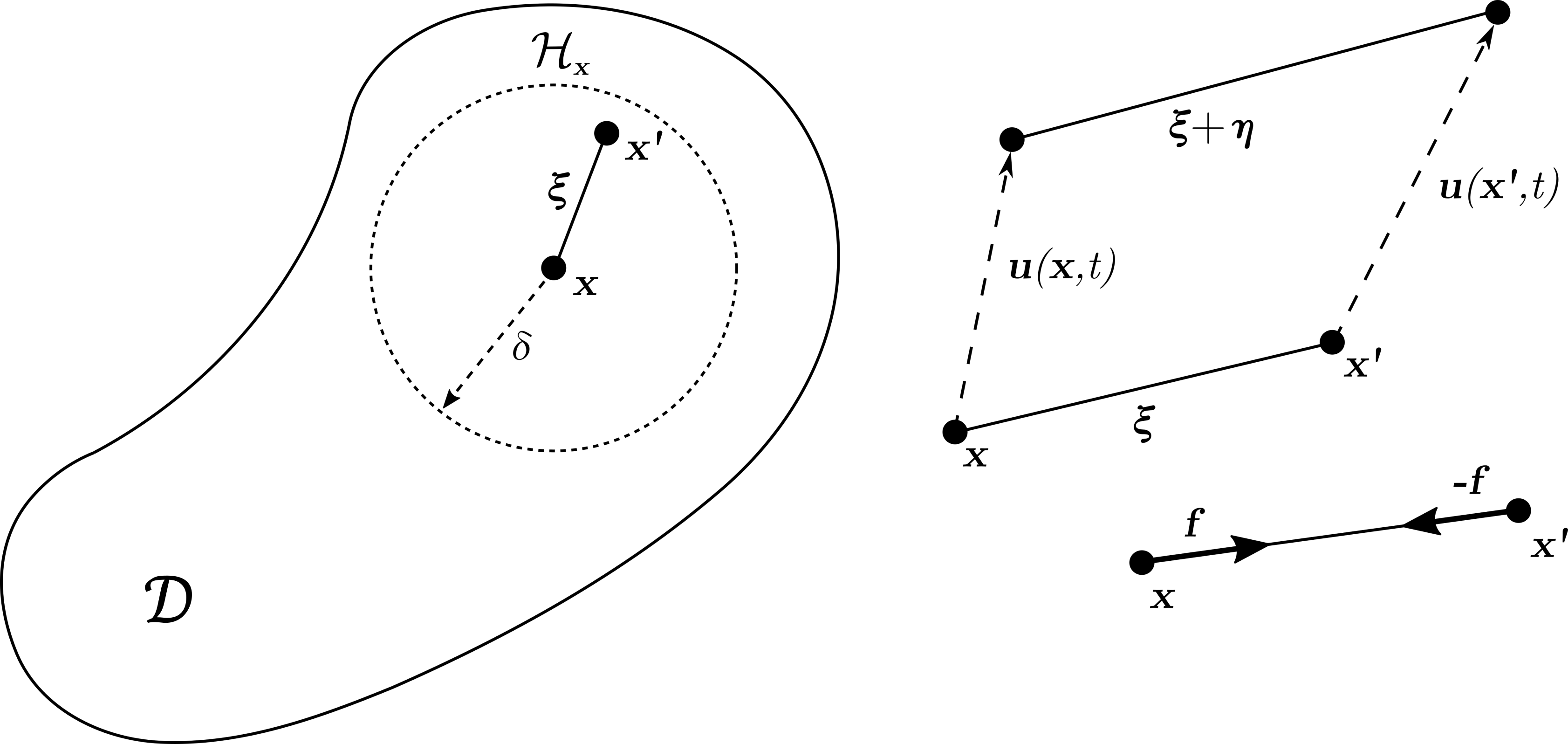}
	\caption{Peridynamic body and kinematics of particle pair and bond-based pairwise force function.}
	\label{fig:particle_pair}
\end{figure}

\subsection{Peridynamic continuum model}\label{sec:peridynamics_continuum} 
\noindent Our starting point, which is general to all formulations of peridynamic models, is the continuum description. A deformable body (see Fig.\ref{fig:particle_pair}) occupies the domain $\mathcal D \subset \mathbb R^d$ (for $d = 2$ or $3$). The deformation (displacement) at some position ${\bm x} \in \mathcal D$ is described by the vector valued function $\bm{u}({\bm x}, t) : \mathcal D \rightarrow \mathbb R^3$, for some time $t \in \mathbb R^+$. The relative distance between two points ${\bm x}$ and ${\bm x}' \in \mathcal D$ in the undeformed and deformed configurations are denoted as
$$
{\bm \xi} = {\bm x}' - {\bm x} \quad \mbox{and} \quad {\bm \eta} = {\bm u}({\bm x}') - {\bm u}({\bm x}).
$$
The acceleration of the body at ${\bm x} \in \mathcal D$ is described by the peridynamic equation of motion
\begin{equation}\label{eq:equation_of_motion}
\rho({\bm x})\ddot{\bm{u}}(\bm{x}, t) = \int_{\mathcal{H}_{x}} \bm{f}(\bm{\eta}, \bm{\xi})\; dV_{\bm{x}'} + \bm{b}(\bm{x}, t).
\end{equation}
where $\rho({\bm x})$ is the material density, $f(\bm{\eta}, \bm{\xi})$ is a pair-wise force function (the force vector per unit volume squared that the body at ${\bm x}'$ exerts on ${\bm x}$)  and ${\bm b}({\bm x}, t)$ denotes a body force. Perhaps most importantly in peridynamic theory, $\mathcal{H}_{x}$ defines the horizon, a subdomain of the undeformed space $\mathcal H_{x} \subset \mathcal D$, over which forces within the material interact. This finite length scale over which forces in a material interact means that peridynamics defines a {\em non-local continuum model} of the material.

\smallskip
The constitutive behaviour of the material is encapsulated within the force function ${\bf f}$. Silling \cite{Silling2005} calculated the pairwise force function for an elastic material, defined in terms of a scalar \emph{micropotential} $w$, such that
\begin{equation}\label{eq:scalar_micropotential}
\bm{f}(\bm{\eta}, \bm{\xi}) = \frac{\partial w}{\partial \eta} (\bm{\eta}, \bm{\xi})  \quad \forall \bm{\eta}, \bm{\xi}.
\end{equation}
This representation leads to a pairwise force function of the form
\begin{equation}
    \bm{f} = \frac{\bm{\eta} + \bm{\xi}}{\|\bm{\eta} + \bm{\xi}\|} f(\|\bm{\eta} + \bm{\xi}\|, \bm{\xi})
\end{equation}
where $f$ is an associated scalar-valued function that contains the constitutive model.

\subsection{Constitutive model}\label{constitutive}
To define the constitutive model, the micropotential function for linear elastic materials is defined as
\begin{equation}\label{eq:microelastic}
w(\bm{\eta}, \bm{\xi}) := \frac{c(\bm{\|\xi\|})s^2 \bm{\xi}}{2}, \quad \mbox{where} \quad s = \frac{\|\bm{\eta} + \bm{\xi}\| - \|\bm{\xi}\|}{\|\bm{\xi}\|}.
\end{equation}
Here, $s$ is the stretch between two points in the material, while $c(\bm{\|\xi\|})$ is the micro-modulus function and refers to the elastic bond stiffness between two points.
For the purposes of this paper, we use constitutive laws that have been derived for a constant micro-modulus function. The use of a non-constant micro-modulus function may provide a number of benefits, such as reduced surface effects \cite{MBobaru2011} and improved numerical convergence \cite{Hu2010, Seleson2014}. We have chosen to implement two micro-modulus functions in \texttt{PeriPy}: constant and conical, which is explored by Ha and Bobaru \cite{Ha2010}. The code can be easily extended to other formulations of $c(\cdot)$.\smallskip

% \begin{equation}\label{eqn:conical_form}
% c(\bm{\|\xi\|}) =c_{1}\left(\frac{\delta-
% \|\xi\|}{\delta}\right)
% \end{equation}

% \begin{figure}
% 	\centering    
% 	\includegraphics[width=0.85\textwidth]{conical.png}
% 	\caption{\mynote{Placeholder figure for the micromodulus functions}
% 	Constant (left) and conical (right) micromodulus functions}
% 	\label{fig:micromodulus}
% \end{figure}
In this paper, we consider examples using the \emph{prototype micro-elastic brittle} (PMB) model \cite{Silling2005}, which relates the bond stretch $s$ to a scalar-valued force projected in the direction of the bond,
\begin{equation}\label{eq:simpleBond}
f = c s \mu.  = \left( \frac{18 K}{\pi \delta^4} \right) s \mu.
\end{equation}
Here, $K$ defines the material bulk modulus, horizon $\delta$, stretch $s$ (see \eqref{eq:microelastic}) and a damage parameter $\mu$, which controls the onset of brittle failure. In the undamaged/linear regime, this expression was derived by equating the strain energy density of a peridynamic body under uniform dilation with classical elasticity theory \cite{Silling2005}. For a pure shear state, there is also an equivalent strain energy density, and a second expression for $c$ can be determined. Equating these two expressions for $c$, it is found that, for the simple PMB peridynamic formulation, the Poisson's ratio $\nu$ is limited to $1/4$ in the general 3D case for plain strain and $1/3$ \cite{Trageser2020} in 2D plane stress conditions. This restriction on the Poisson's ratio can and has been addressed by introducing state-based peridynamic theory, which has been considered extensively by Silling et al. \cite{Silling2007}. However, it is not discussed further in this paper or used within the examples since, while the initial formulation of the ordinary state-based model is more involved, the resulting computational implementation is equivalent, albeit with additional parameters to define a material response (see Section \ref{sec:StateBased}).

\begin{figure}[h!]
	\centering
	\includegraphics[width=0.4 \linewidth]{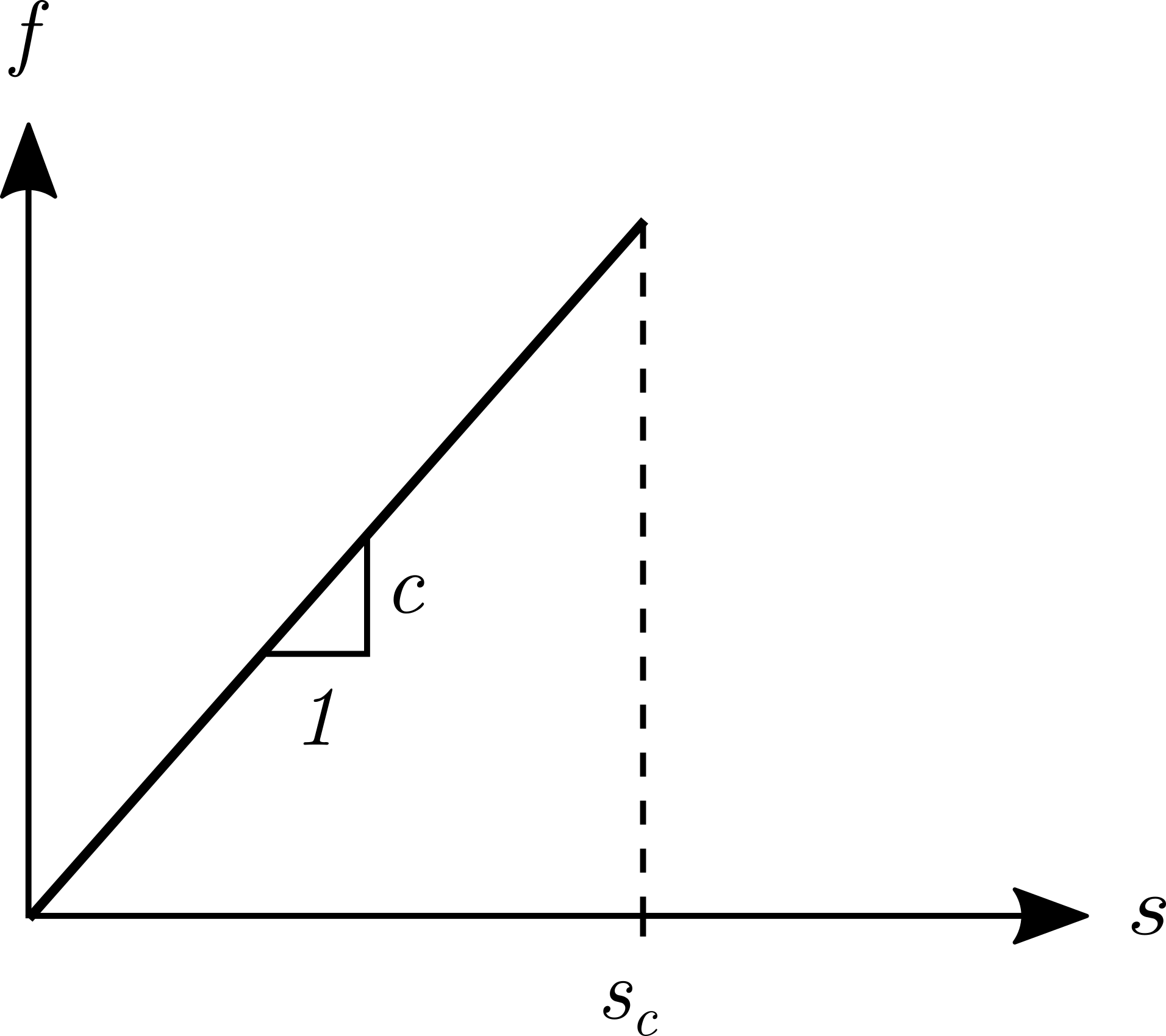}
	\caption{The PMB model, also known as the linear damage model.}
    \label{fig:linear_damage_model}
\end{figure}

\smallskip
As introduced in \eqref{eq:simpleBond}, brittle failure is captured in the constitutive model by introducing a history-dependent scalar-valued (boolean) function
\begin{equation}\label{eqn:mu}
\mu =
\begin{cases}
1 \quad \mbox{for} \quad s < s_{c} \; \mbox{and} \; \max_{t' \in [0,t]} s(t') < s_{c}  \\
0 \quad \mbox{otherwise}
\end{cases}.
\end{equation}
This (non-smooth) parameter allows bonds to irreversibly break when they are stretched beyond a predefined limit, defined by the {\em critical strain} of the bond,
\begin{equation}\label{eqn:critical_stretch}
s_c =
\sqrt{\frac{5G_F}{6E\delta}} \quad \mbox{(3D)}, \quad s_c =
\sqrt{\frac{4 \pi G_F}{9E\delta}} \quad \mbox{(plane stress)} \quad \mbox{and} \quad
s_c = \sqrt{\frac{5 \pi G_F}{12E\delta}} \quad \mbox{(plane strain)}.
\end{equation}
where $E = 3K(1 - 2\nu)$ is Young's modulus, and $G_F$ the fracture energy release rate of the material. Silling and Askari \cite{Silling2005} derived this value by separating a peridynamic body into two halves and thereby breaking all the bonds that initially crossed the fracture surface. By determining the energy required to break a single bond, the energy per unit fracture area for complete separation of a peridynamic body can be determined. The energy release rate $G_F$ is a material parameter and can be determined experimentally. The PMB model is shown in Fig.~\ref{fig:linear_damage_model}.

\medskip\noindent
{\em Remark:} Although in this paper we limit the examples to modelling quasi-brittle homogeneous isotropic materials, \texttt{PeriPy} can model heterogeneous materials, such as composites. Mehrmashhadi et al. \cite{Mehrmashhadi2019} randomly assigned bonds as matrix, fibre or interface bonds so that the volume fraction matched the average volume fraction of peridynamic nodes, thereby modelling a composite material that is homogeneous on length scales larger than the horizon. This is possible in \texttt{PeriPy} by defining a bond classification function as a function of the points $\bm{x}$ and $\bm{x}'$. The bond stiffness and other damage model parameters for the matrix, fibre or interface bonds are passed at model instantiation or, alternatively, at the time of simulation, which allows for outer-loop black-box optimisation of constitutive parameters with respect to some objective function, as demonstrated in Section~\ref{concrete_beams}. Since the relative angle between a bond and the fibre orientation represents the fibre directions, peridynamic theory has an inherent advantage in modelling fibre-reinforced composite laminates. Xu et al. \cite{Xu2008} classified the bonds as either fibre or matrix bonds to reproduce the anisotropy of a fibre-reinforced composite lamina. \texttt{PeriPy} is also easily capable of modelling anisotropic materials, which may be modelled through a dependence of $\bm{f}$ on the initial direction of the bond $\bm{\xi}$. Due to the inherent anisotropy of a fibre-reinforced composite, Hu et al. \cite{Hu2014} assigned distinct micro-elastic moduli to the bonds to describe the anisotropy. The bonds were classified as either longitudinal or transverse bonds. In their work, they used a micromodulus function to introduce the dependence of the interaction strength on the bond length. Micromodulus functions are included in \texttt{PeriPy}, and the code is easily extended for users to implement their own micromodulus function, such as the one used by Hu et al. The design choices and features of \texttt{PeriPy} are discussed in more detail in Section~\ref{design_overview}.\smallskip

The continuum level \emph{damage} in a peridynamic body can be measured by introducing the local damage $\phi(\bm{x}, t)$
\begin{equation}
\phi(\bm{x}, t) = 1 - \frac{\int_{\mathcal{H}_{{\bf x}}} \mu (\bm{x}, t)dV}{\int_{\mathcal{H}_{{\bf x}}}dV}.
\end{equation}
This defines the ratio of broken bonds over original bonds for all points in the horizon of ${\bm x}$. Therefore, $0 \leq \phi \leq 1$, with 0 representing virgin material, and 1 represents complete disconnection of a point from all of the points with which it was initially connected. To completely separate the two halves of the body across the fracture surface requires breaking all the bonds that initially connected points in the opposite
halves and will result in a damage of $\phi \approx 0.5$ at those points on the fracture surface.\smallskip

\subsection{Spatial disretisation}\label{discrete_peridynamics}
As before, let a body in its initial configuration occupy a domain $\mathcal{D} \subset {\rm I\!R}^d$. Different disretisation approaches can be used to solve a problem numerically in peridynamics  \cite{Emmrich2007}. Computationally, using the mesh-free approach of Silling and Askari \cite{Silling2005}, the material can be represented by nodes (or {\em particles}) in the index set $\mathcal{G} := \{ i \; | \; \bm{x}_{i} \in \mathcal{D} \}$. Here, the total number of nodes is denoted by $n = |\mathcal{G}|$, the initial coordinates by $\bm{x}_i \in \mathcal D$, the displacement vector by $\bm{u}_i (t) \in [0,T] \times {\rm I\!R}^d$ and associated lumped volumes by $dV_i$ for that node. Furthermore, a given node is subject to an externally applied force density $\bm{b}_i \in  {\rm I\!R}^d$.\smallskip

As mentioned above, the peridynamic formulation is a \emph{non-local} model. The $i^{th}$ node interacts with the all nodes within the index set
\begin{equation}\label{eq:family}
\mathcal H_i := \{ j \; | \; \| \bm{x}_{i} - \bm{x}_{j} \|\leq \delta \},
\end{equation}
which is called the \emph{family} of node $\bm{x}_{i}$. The user-defined parameter $\delta \geq 0$ is, as before, the \emph{horizon}, which describes a characteristic length scale specific to the mechanics of the material.\smallskip

The discretised peridynamics equation of motion approximates the integral in Eq. \eqref{eq:equation_of_motion} with the sum
\begin{equation}\label{eq:discrete_equation_of_motion}
\rho\ddot{\bm{u}}_{i}(\bm{x}, t) = \sum_{j \in \mathcal{H}_{i}} \bm{f}(\bm{\eta}_{ij}, \bm{\xi}_{ij})\beta_{ij}dV_j + \bm{b}_{i}(\bm{x}, t).
\end{equation}
where a partial volume correction $\beta_{ij}$ is introduced to improve the accuracy of the spatial integration. For simplicity, this work does not use any partial volume correction algorithms, but \texttt{PeriPy} implements the partial volume algorithm proposed by Hu {\em et al.} \cite{Hu2010}. \citet{Seleson2014} provides a detailed analysis of the various methods available for partial volume correction algorithms and concludes that this algorithm outperforms approximation of the partial volumes, both with their full volume ($\beta = 1$) and with the algorithm that is commonly used in peridynamics software (e.g. PDLAMMPS \cite{PDLAMMPS}).

\smallskip
An energy dissipation term, known as dynamic relaxation, can be added to \eqref{eq:discrete_equation_of_motion}, so the solution will converge to a steady state solution \cite{Kilic2008}. The method relies on the introduction of an artificial damping term in the equation of motion,
\begin{equation}\label{eq:equation_of_motion_damped}
\rho\ddot{\bm{u}}_{i} + \eta\dot{\bm{u}}_{i} = \sum_{j \in \mathcal{H}_{i}} \bm{f}(\bm{\eta}_{ij}, \bm{\xi}_{ij})\beta_{ij}dV_j + \bm{b}(\bm{x}, t) := \bm{F}_i.
\end{equation}
where the resultant force on the node $i$, $\bm{F}_i$ has been defined for brevity.

\subsection{Time integration}\label{time_integration}
Simulating the evolution of the model from its initial conditions requires a numerical time integration scheme. The choice is between \emph{explicit} and \emph{implicit} schemes \cite{Littlewood2013}. Explicit schemes are most commonly used in peridynamics solvers because they are more suited to parallel computation, as no solution of a system of equations is required. The specific method chosen was the velocity-Verlet method. Velocity-Verlet leads to improvements in numerical accuracy in node dynamics simulations \cite{Groot1997}, is simple to implement and has been widely used in other implementations of peridynamics solvers. Velocity-Verlet is a second-order symplectic integrator that has minimal computational memory requirements, which is important for GPUs since they are inherently memory constrained by the amount of global memory, or VRAM. The velocity-Verlet scheme can be written as follows:
\begin{equation}\label{eq:Velocity_Verlet1}
\bm{u}_{i}(t + \Delta t) = \bm{u}_{i}(t) + \Delta t \bm{\dot{u}}_{i}(t) + \frac{\Delta t^{2}}{2} \bm{\ddot{u}}_{i}(t)
\end{equation}
followed by
\begin{equation}\label{eq:Velocity_Verlet2}
\bm{\dot{u}}_{i}(t+ \Delta t) = \bm{\dot{u}}_{i}(t) + \frac{\Delta t}{2}\left(\bm{\ddot{u}}_{i}(t) + \bm{\ddot{u}}_{i}(t + \Delta t)\right).
\end{equation}
Here, $\bm{\ddot{u}}_{i}(t)$ is the acceleration term evaluated using the equation of motion \eqref{eq:discrete_equation_of_motion}. The velocity-Verlet algorithm does not provide a prescription for including velocity dependent forces as a result of dynamic relaxation. To extend this algorithm to include velocity-dependent forces, we follow Groot and Warren (1997) \cite{Groot1997}, where the equation of (damped) motion \eqref{eq:equation_of_motion_damped} is written as follows:
\begin{equation}\label{eq:Velocity_Verlet3}
\bm{\ddot{u}}_{i}(t + \Delta t) = \frac{\bm{F}_{i}(t)  - \eta \bm{\dot{u}}_{i}(t + \frac{\Delta t}{2})}{\rho},
\end{equation}
where the half step velocity is $\bm{\dot{u}}_{i}(t + \frac{\Delta t}{2}) = \bm{\dot{u}}_{i}(t) + \frac{\Delta t}{2} \bm{\ddot{u}}_{i}(t)$.

\subsection{Surface correction factors}
The derivation of the peridynamic bond stiffness $c$ is based on the assumption that the horizon of a material point is contained within the bulk $\mathcal{H}_{i} \in \mathcal D$. This is not true for nodes within a distance $\delta$ from the body edge, which do not possess a full family. Consequently, the stiffness of the nodes contained within the bulk and nodes on the edge of a body differs, resulting in significant error between peridynamic solutions and classical analytical methods. This is called the peridynamic surface effect. Le and Bobaru \cite{Le2018} presented a detailed comparison of the most common surface effect correction techniques. They found that the Volume method, proposed by Bobaru et al. \cite{Bobaru2017}, is one of the most effective techniques for correcting surface effects. The method ensures that nodes under homogeneous deformation have the same strain energy density values when located near the surface or in the bulk. The micromodulus of a bond $\xi_{ij}$ is corrected as $c_{ij,\emph{corrected}} = \lambda_{ij} c_{ij}$ where
\begin{equation}
\lambda_{ij} = \frac{2V_0}{V_i + V_j}
\end{equation}

\section{Implementation of \texttt{PeriPy} in python and \texttt{OpenCL}}\label{implementation}
It has been proposed that, if a peridynamic code has few dependencies, is easy to use and, above all, is scalable and sufficiently fast for real-world engineering problems, then the user base of bond-based peridynamics will increase, and new research in bond-based peridynamics can utilise a solver with a known, robust standard. To that end, we have developed an open-source GPU (Graphics Processing Unit) code that uses \texttt{OpenCL} \cite{opencl_website} \cite{opencl_spec} for GPU acceleration and a python front-end for user inputs. Details of installation (via \texttt{pip}) and documentation can be found in \ref{Installation}. The GPU communicates with the \textit{host} device (a CPU) with the \texttt{OpenCL} API. The front-end of the code is written in python in order to make it accessible for research, and as a result the python API, \texttt{pyopencl} \cite{pyopencl}, was chosen.

\smallskip
Preliminary testing showed that the calculation of bond forces and their reduction to a single node force were the most expensive components for solving any peridynamics problem numerically. Testing showed that these two steps contribute to 93–98\% of the run time over a range of test setups. This is because of the need to calculate the bond force density of each bond for each node family in Eq. \eqref{eq:discrete_equation_of_motion}. Our implementation therefore targets the acceleration for these two critical steps in the peridynamic solver. 

\smallskip
In a serial peridynamics solver, a nested for-loop over all nodes and families is used to obtain the force density contributions of each bond $\bm{f}_{ij}$, with each of the elements of this nested for-loop calculated in series. In fact, the bond forces $\bm{f}_{ij}$ can be readily calculated in parallel, as the calculation of the contribution to force density of one bond is independent of another. Therefore, calculation of the bond forces are ideal for GPU acceleration since each bond calculation can be done on one of the many concurrent threads in a GPU. Peridynamics is particularly suited to parallelisation when compared to other particle methods, such as Smoothed Particle Hydrodynamics (SPH), because the family $\mathcal{H}_i$ remains unchanged throughout the simulation, and thus a neighbour search over nodes only needs to be performed once for any given model, and for each subsequent simulation, the family can be read from a file stored to disk.\smallskip

\smallskip
A function that is written in \texttt{OpenCL} (using C) that replaces loops with functions executing at each point in a problem domain is called a compute kernel or \emph{kernel}. This is the central idea behind \texttt{OpenCL}. Mossaiby et al. (2017) \cite{Mossaiby2017} demonstrated that, by replacing the outer loop over nodes with a kernel for each node executing in parallel, a speed increase of a factor of $40$ to $90$ times over optimised sequential \texttt{C++} code and an increase of $3$ to $6$ times over \texttt{OpenMP} parallel code can be achieved. The current work improves on the work of Mossaiby et al. \cite{Mossaiby2017} by pairing a parallel computation of each bond force and a binary parallel reduction of the bond forces into the nodal force density by exploiting the \emph{execution} and \emph{memory} models of \texttt{OpenCL}, respectively. We elaborate on this step below and give a brief introduction to \texttt{OpenCL} with an example.

\subsection{Introduction to \texttt{OpenCL} with binary parallel reduction example}\label{data_structure}

We will now introduce the language and concepts of \texttt{OpenCL} by way of explaining one of the optimisations used to speed up the code in the current work; binary parallel reduction, which is a common and important data parallel primitive \cite{Harris2007}.

\subsubsection{Execution model}

\begin{figure}[ht]
	\centering    
	\includegraphics[width=9cm]{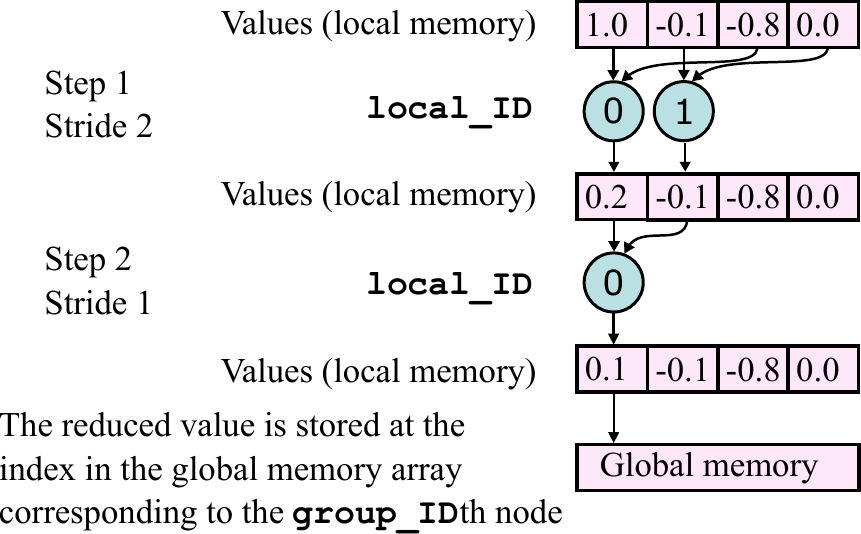}
	\caption{Schematic of binary parallel reduction of a force in one Cartesian direction for one work-group representing the ith of $n$ nodes, for $N=4$.}
	\label{fig:reduction}
\end{figure}

In mesh-free peridynamics, the bond force density contributions are summed to calculate the resultant force on each particle, as shown on the right hand side of Eq.~\ref{eq:discrete_equation_of_motion}. In \texttt{PeriPy}, the bond force density contributions for each node for each Cartesian direction are stored in an $N \times 1$ \texttt{C} array and summed to a Cartesian resultant force vector using a binary parallel reduction algorithm. A pictorial representation of this algorithm for a single node and Cartesian direction and $N=4$ is shown in Fig.~\ref{fig:reduction}. The idea of this algorithm is to use sequential addressing \cite{Harris2007} to perform an addition on two values that are a \emph{stride} apart and propagate this result downwards, at each step reducing the stride by a factor of two, so that the final result is inserted into the first space of a local memory cache. The pseudocode for binary parallel reduction is shown in Algorithm~\ref{alg:reduction} in the Appendix, and is executed using the \texttt{OpenCL} kernel shown in Listing~\ref{lst:kernel}, also in the Appendix.

The \texttt{OpenCL} execution model \cite{opencl_spec} describes how kernels execute. When a kernel is submitted for execution by the host (as defined in Section~\ref{implementation}), an index space is defined. An instance of the kernel, called a \emph{work-item}, executes for each point in this index space. The work-item can therefore be identified by its point in the index space, called the \texttt{global\_id}. Returning to the binary parallel reduction example in Listing~\ref{lst:kernel}, to create the index space, each bond in the sparse two-dimensional index set $\{\mathcal{H}_1, ... , \mathcal{H}_n\}$ is assigned a $\texttt{global\_id} \in [0, \texttt{global\_size})$, where $\texttt{global\_size} = nN$.

\smallskip
In \texttt{OpenCL}, work-items are organised into \emph{work-groups}. The work-groups provide a coarse decomposition of the index space. The \texttt{local\_size} is the number of work-items per work-group. A work-group is identified by its point in this coarse index space, called the \texttt{group\_id}. Thus, each and every node may be assigned a \texttt{group\_id} if each work-group contains $\texttt{local\_size}=N$ work-items, as is the case in \texttt{PeriPy}. Work-items are assigned a unique $\texttt{local\_id} \in [0, N)$ within a work-group such that $\texttt{global\_id} = \texttt{group\_id} \times \texttt{local\_size} + \texttt{local\_id}$. Calculations on the work-item are then for the \texttt{local\_id}th bond of the \texttt{group\_id}th node. The result of executing Listing~\ref{lst:kernel} means that $N$ work-items (for each work-group) execute in parallel to reduce the bond forces to the node force of a single node which is stored in \texttt{local\_cache[0]}. Of course, there must be a \texttt{local\_cache} for each work-group that stores the bond forces for the work-group's node. This is explained by the \texttt{OpenCL} memory model \cite{opencl_spec} in the next section. $N$ must clearly be a power of two so that a binary parallel reduction can be performed. So, $N = \min_{k}(2^k) \geq \max_{i}(|\mathcal{H}_i|)$ is the minimum power of two, which is larger than (or equal to) the maximum number of neighbours of any one node, such that each bond is assigned a \texttt{global\_ID} and the constants that are associated with each bond. For example, the \texttt{stiffness\_corrections} factors can be stored in a dense \texttt{global\_size} \texttt{C} array. Finally, the first work-item in the work-group, \texttt{local\_id = 0} is used to assign this value in the \texttt{body\_force} array and the reduction is complete. It is important to note that not only are each work-item in a work-group executed in parallel, but many work-groups can be executed in parallel, also. See that this means we are using the execution model to parallelise over bonds as well as nodes.

\smallskip
As well as using the execution model to perform the binary parallel reduction algorithm, we use the execution model to calculate the other computationally expensive part, the bond-forces. Each work-item is responsible for calculating the bond force of one bond, which is presented in the pseudocode of Algorithm~\ref{alg:bond_index} in the Appendix. This is in contrast to the work of Mossaiby et al \cite{Mossaiby2017}, who assign an index space only over nodes, which is presented in pseudocode in Algorithm~\ref{alg:node_index}, also in the Appendix. In the kernel for Algorithm~\ref{alg:bond_index}, work-items and work-groups are assigned unique \texttt{global\_id} and \texttt{group\_id} respectively in exactly the same way as in Listing~\ref{lst:kernel}. The index of the child node is $j = \texttt{nlist}[\texttt{global\_id}]$, where \texttt{nlist} is the neighbour list and is an $n \times N$ \texttt{C} array where the $i$th row contains $\mathcal{H}_i$. The rows of \texttt{nlist} are padded with values of $-1$, where $|\mathcal{H}_i| < N$. Note that node indices are taken from the set of non-negative integers, so $-1$ is a safe choice for denoting a broken bond. The \texttt{global\_id}th bond is broken by setting the value of $\texttt{nlist}[\texttt{global\_id}]  = -1$ and the bond force $f_{ij} = 0$. Using Algorithm~\ref{alg:bond_index}, it is possible to achieve particular optimisations that result in a significant speedup in the peridynamics code, as explained in Sections~\ref{kernel_fusing}-\ref{thread_divergence}.

\begin{figure}[H]
	\begin{subfigure}{0.5\textwidth}
		\includegraphics[width=9.0cm]{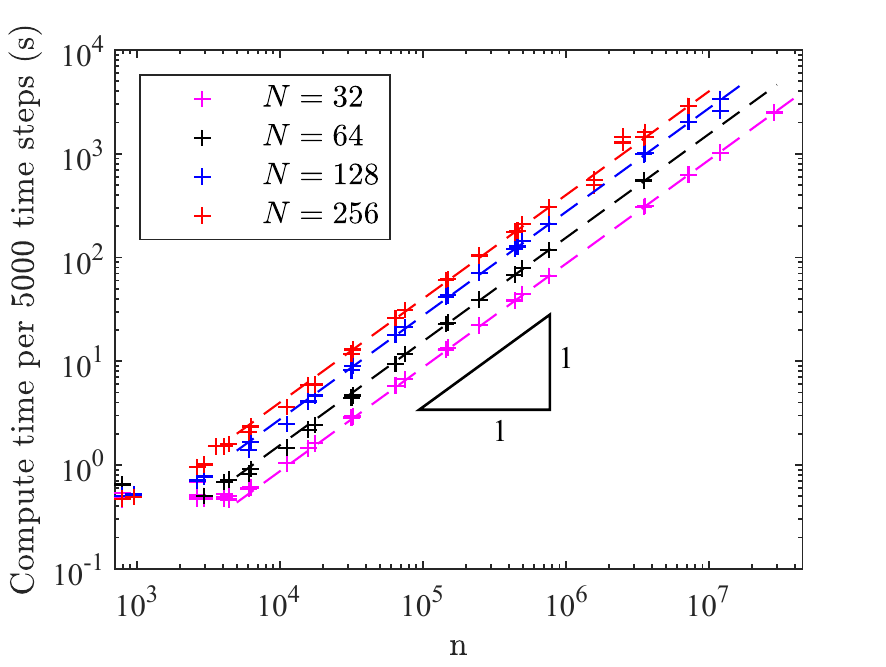}
		\caption{Without stiffness correction factors.}
		\label{fig:profile_no_regular}
	\end{subfigure}
	\begin{subfigure}{0.49\textwidth}
		\includegraphics[width=9.0cm]{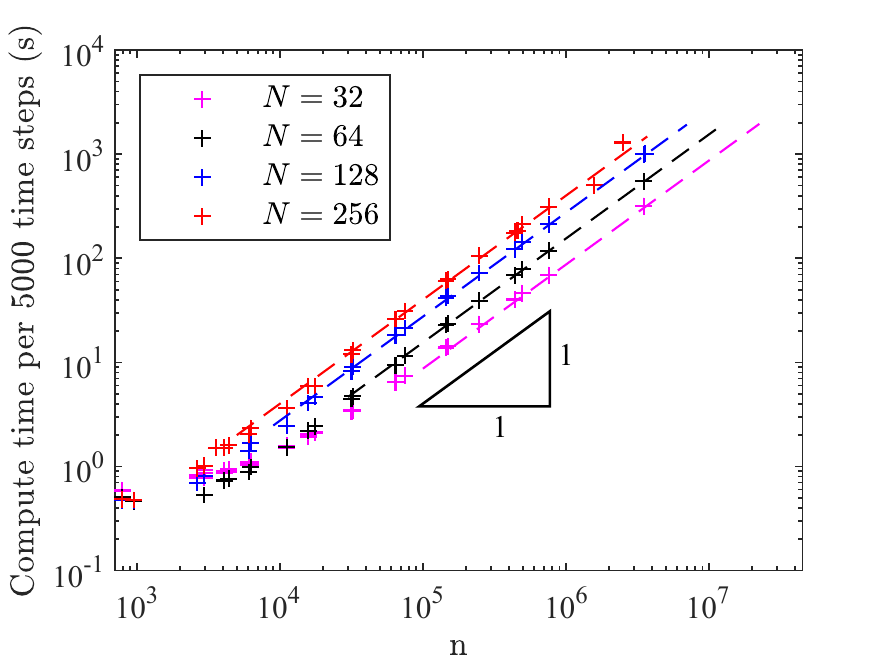}
		\caption{With stiffness correction factors}
		\label{fig:profile_s_regular}
	\end{subfigure}
	\hfill
	\caption{\texttt{PeriPy} performance across problem size for 5000 time steps with a regular mesh. $n$ is the number of nodes and $N \geq \max_{i}(|\mathcal{H}_i|)$ is the maximum number of nodes in the non-local family. A number of problem sizes were used to demonstrate the linear scaling performance of \texttt{PeriPy} for problems with a large number of nodes. The benchmark tests were run on an NVIDIA 2080Ti GPU. The particular geometries and boundary conditions do not affect performance, and simple ones were applied.}
	\label{fig:profile_regular}
\end{figure}

\smallskip
In \texttt{OpenCL}, the $\texttt{global\_size} = nN$ must be divisible by the $\texttt{local\_size} = N$ because work-items must be divided between work-groups. Recall that there is a one-to-one mapping between \texttt{group\_id}s and nodes, and each work-group is responsible for calculating the resultant node force of a single node. Work-groups are executed in parallel on a device's compute units. There is no way to synchronise the timings of work-groups since, in an application, there are usually more work-groups than hardware to execute them concurrently. As a result, they will be queued to run as compute units become available. Due to the limited number of compute units on any GPU, we can expect the performance of \texttt{PeriPy} to scale linearly with the number of nodes. At smaller problem sizes, the overhead due to memory transfer between the GPU and controlling CPU dominates, while at larger problem sizes the time required for actual computation comprises most of the total wallclock time. This is seen to be the case in Fig.~\ref{fig:profile_regular}, which shows the performance of \texttt{PeriPy} for different values of $N$. All of the benchmarking tests in Fig.~\ref{fig:profile_regular} and Fig.~\ref{fig:profile_irregular}-\ref{fig:implementations_256} were run on an NVIDIA 2080Ti GPU for various problem sizes $n$. The benchmarking setup is detailed in Section~\ref{benchmarking}. Fig.~\ref{fig:profile_regular} shows tests for beams that have been discretised with a regular cubic tensor grid, consistent with the majority of the existing literature. The beams were put under some basic boundary conditions, and the tests were displacement controlled, although this does not affect the simulation time. The stable time-step condition derived by Silling and Askari \cite{Silling2005} was used.

Note that $N$ can be chosen independently of $n$. This is because the horizon radius $\delta$ is selected so that it has a value $\delta > m\Delta x$, where $m$ is a small positive constant (usually chosen to be $\pi$) \cite{Gerstle2010} and $\Delta x$ is the mesh spacing. The choices of $m$ and $\delta$ are discussed in \citet{Bobaru2012} and \citet{Henke2014}. In the limit of the horizon $\delta$ approaching zero, the peridynamic solution should converge to the elastic classical continuum mechanics elasticity solution. Bobaru and Hu \cite{Bobaru2012} and Zhao et al. \cite{Zhao2020} note that the size of the horizon should be defined by the smallest geometrical feature; see also \citet{Silling2005} and \citet{Silling2019}.

\subsubsection{Memory model}
\begin{figure}[ht]
	\centering    
	\includegraphics[width=9cm]{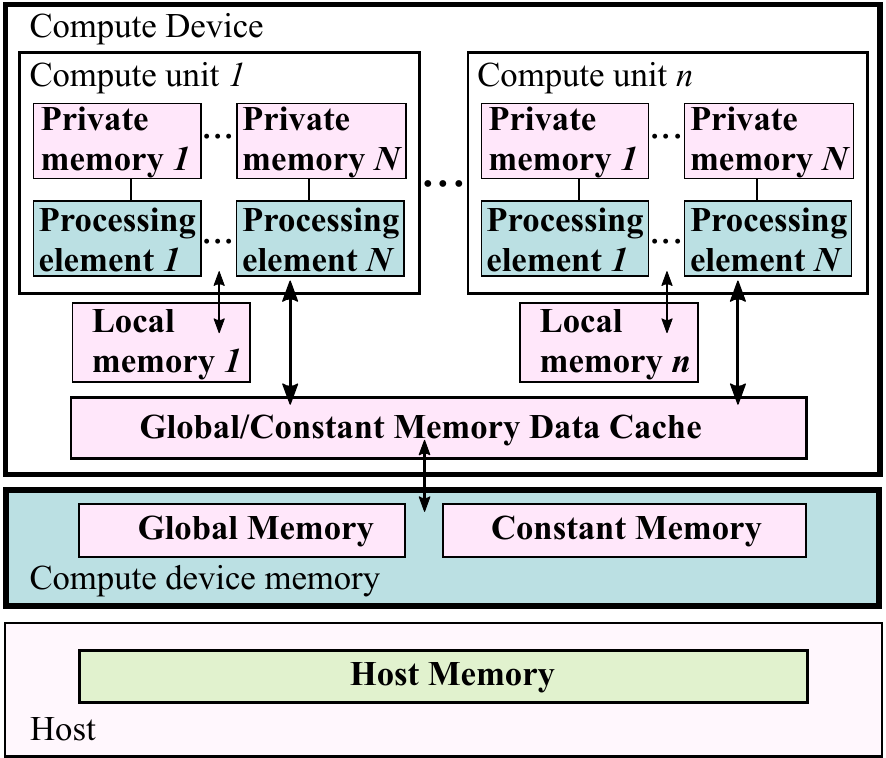}
	\caption[node-pair]{The named address spaces exposed in an \texttt{OpenCL} Platform. Global and Constant memories are shared between the one or more (not shown) devices within a context, while local and private memories are associated with a single device. Each device may include an optional cache to support efficient access to their view of the global and constant address spaces. This figure is adapted from the \texttt{OpenCL} specification \cite{opencl_spec}.}
	\label{fig:memory_model}
\end{figure}

The memory regions of an \texttt{OpenCL} device and their relationship to the \texttt{OpenCL} platform model \cite{opencl_spec} are summarised in Fig.~\ref{fig:memory_model}. The most basic building block of the \texttt{OpenCL} platform model is the processing element: {\em a virtual scalar processor, one of the many concurrent 'threads' on the device}. A work-item may execute on one or more processing elements. A compute unit is composed of one or more processing elements and local memory. An \texttt{OpenCL} device has one or more compute units, and a work-group executes on a single compute unit.

\smallskip
Compute device memory is one or more memories attached to the compute device that are directly available to kernels executing on \texttt{OpenCL} devices. Compute device memory consists of four name address spaces or memory regions:

\begin{itemize}
    \item \emph{Global Memory}: permits read/write access to all work-items in all work-groups running on any device within a context. Work-items can read from or write to any element of a memory object. Reads and writes to global memory may be cached depending on the capabilities of the device.
    \item \emph{Constant Memory}: A region of global memory that remains constant during the execution of a kernel instance. The host allocates and initialises memory objects placed into constant memory.
    \item \emph{Local Memory}: A memory region local to a work-group. This memory region can be used to allocate variables that are shared by all work-items in that work-group.
    \item \emph{Private Memory}: A region of memory private to a work-item. Variables defined in one work-item's private memory are not visible to another work-item.
\end{itemize}

Local and private memories are always associated with a particular compute unit and processing element, respectively. Local and private memories are much faster than global memory. Therefore, it is imperative to exploit the different memory regions in GPU-accelerated programming. Consider again the binary parallel reduction example in Listing~\ref{lst:kernel}. The local memory cache was used to store the three (one for each Cartesian direction) $N \times 1$ \texttt{C} arrays containing the bond forces. The summation is performed using the faster bandwidth of the local memory. The bond forces are never stored in the slower-to-access global memory.

\smallskip
The global and constant memories are shared memories between all compute devices, compute units and processing elements. In the binary parallel reduction algorithm, the \texttt{body\_force} array is stored in global memory. In the bond force algorithm, the \texttt{stiffness\_corrections} are constant and so stored in the constant memory. In \texttt{PeriPy}, a single compute device is used, which can be either the GPU or CPU. The four named address spaces available to a device do not overlap. This is a logical relationship, however, and an implementation may choose to let these disjointed named address spaces share physical memory.

\smallskip
On the GPU, the private memory accessible to each work-item is small in size and is only $O(10)$ (32-bit) words per work-item and very fast \cite{hands_on_opencl}. Non-pointer variables are stored in the private memory, such as those used to assign the node indices $i=\texttt{group\_id}$ and $j=\texttt{nlist}[\texttt{global\_id}]$ and relative displacements $\bm{\xi}_{ij}$. Private memory is used to do fast floating-point arithmetic, such as evaluating the bond force. Each work-group contains many work-items that share its local memory. This memory is larger ($O(10)$ kBytes) than the private memory of a work-item and is used in \texttt{PeriPy} to store the bond forces. The largest memory on a GPU is the global memory, which can be in the high-speed GPU memory (VRAM).

\subsection{Hardware considerations}
There are a number of hardware-dependent considerations that users of \texttt{PeriPy} should note. These are:

\smallskip
\textbf{Memory bandwidth:} Memory transfers to CPU are expensive, and thus the whole application is transferred onto the GPU until completion unless intermediate writes are performed.

\smallskip
\textbf{Local memory size:} For application to the vast majority of 3D problems in the literature, the $\texttt{local\_size} = N$ is usually $64$, $128$ or $256$ depending on $m$ and, as mentioned above, is a power of two so that a parallel binary parallel reduction can be performed. This can work well since, for NVIDIA cards, which are multiprocessor units that create, manage, schedule and execute threads in groups of 32 parallel threads called {\em warps} \cite{nvidia_programming_guide}, any \texttt{local\_size} that is a multiple of $32$ should result in good thread utilisation. These are also the recommended \texttt{local\_size} for intel CPUs \cite{intel_programming_guide}, in the event the code is to be run on a CPU. The amount of local memory used by a work-item will be the limiting factor of the number of work-items that can be executed in parallel \cite{opencl_spec}, as each work-group offers a limited amount of shared local memory and registers. As a result, the hardware limit on the value of $N$ is \texttt{CL\_KERNEL\_WORK\_GROUP\_SIZE} for the kernel that calculates bond forces, which is hardware dependent. On modern cards, this number will be well above the number needed for peridynamics simulations, so is not a problem.

\smallskip
\textbf{Global memory size:} The upper limit on the problem size possible for peridynamics simulations on GPUs is governed by the size of the global memory. The total number of bytes of arrays that are allocated space or stored on the GPU must be less than the size of the global memory. The global memory is used to store the state variables, such as displacement $\texttt{u}[i] = \bm{u}_i$ and \texttt{nlist} arrays. If a simulation uses stiffness correction factors, they are stored in an $n \times N$ double \texttt{C} array in the global memory. The constants, such as node coordinates $\texttt{coords}[i] = \bm{x}_i$ and volumes $\texttt{vols}[i] = dV_i$, are stored in a memory region of global memory called the constant memory, which remains constant during the execution of a kernel. The host allocates and initialises memory objects placed into constant memory. Work-items have read-only access to these objects. These values can be used to approximate the maximum problem size $n$ for a given amount of global memory. The upper limits on $n$ for the NVIDIA 2080Ti GPU with 11Gb of VRAM are shown in Table~\ref{table:memory_limitation_11Gb}. These values define the upper bound for the problem size in the benchmarking results in Section~\ref{benchmarking}. The approximate calculations used to determine the maximum problem size in \texttt{PeriPy} are shown in Table~\ref{table:memory_limitation} in the Appendix.

\begin{table}
	\caption{Approximate maximum problem size in millions ($10^6$) of nodes $n$ for a \texttt{PeriPy} simulation on a GPU with 11Gb of VRAM and work-group size $N$.}
	\centering
	\label{table:memory_limitation_11Gb}
	\begin{tabular}{l c c c}
		\toprule
		\text Features used in \texttt{peripy.model} class & $N=64$ & $N=128$ & $N=256$\\ 
		\midrule
		single material, linear damage model, no stiffness corrections & 29.8 & 18.1 & 10.1 \\
		
    	single material, linear damage model, stiffness corrections & 13.0 & 7.0 & 3.7 \\
	
		multiple material, n-linear damage model, no stiffness corrections & 13.0 & 7.0 & 3.7\\
		
		multiple material, n-linear damage model, stiffness corrections & 8.3 & 4.4 & 2.2\\
		\bottomrule
	\end{tabular}
\end{table}

\smallskip
\textbf{Precision:} Single precision performance (in FLOPS) on GPUs is better than double precision performance \cite{Eijkhout2016}. Our tests showed that using single precision with \texttt{PeriPy} is between 6.7 and 8.6 faster than using double precision on our GPU. Now that the memory model has been explained, the optimisations of the current work are demonstrated.

\subsection{Optimisation 1: Kernel fusing}\label{kernel_fusing}
Each kernel launch has some overhead, which includes the allocation of memory for the first launch, copying the values of the input parameters to constant memory, indexing any local or global arrays used within the kernel and calculating intermediate variables, such as the bond stretch $s$. By fusing the kernels for calculation and summation of bond forces, the extra overhead is eliminated and the total run time reduced. It is also possible to fuse the kernel that checks whether a bond has exceeded the critical stretch into the bond force calculations kernel, which avoids the need to calculate the bond stretch $s$ twice as well as the need to store the bond forces in an $n \times N$ \texttt{C} array in global memory, which would be accessed by the separate check-bonds kernel. Since there is a work-item for each bond, there are no race conditions that result from fusing these kernels.

\smallskip
The check for broken bonds is applied directly after a displacement update and before the calculation of bond forces. It is important to ensure that all of the bond forces are updated for a given time-step before the time integration is performed to avoid race conditions. Therefore, it is not possible to fuse the time integration kernel with the bond force kernel. Splitting the time integration and bond force kernels guarantees that bond forces are calculated with nodal displacements from the same time step, and vice versa. The time integration kernel for velocity-Verlet is shown in Algorithm~\ref{alg:velocity_verlet} in the Appendix.

\subsection{Optimisation 2: Parallel reduction of bond forces onto nodal forces}\label{parallel_binary_reduction}

The implementation of binary parallel reduction was explained in Section~\ref{data_structure}. The reduction is needed to reduce the bond forces to the node force. The binary parallel reduction facilitates the parallelisation over bonds as opposed to nodes, which leads to the most significant optimisation of the current work, minimising thread divergence.

\subsection{Optimisation 3: Thread divergence}\label{thread_divergence}
The most significant performance consideration relates to the execution of threads or work-items \cite{Niemeyer2014}. Recall that work-items are organised into work-groups. Work-groups can be executed by the GPU in any order, but work-groups are not necessarily execution units themselves. Instead, work-groups are partitioned into \emph{warps} for execution. In the current generation of NVIDIA devices, each warp consists of 32 threads. If a work-group consists of more than 32 work-items, the work-group is partitioned into multiple warps based on the work-item ID (e.g. \texttt{local\_id}). All work-items in a warp should follow the same instruction path. Otherwise, work-items execute different instructions, which is called thread divergence and reduces performance significantly. To avoid thread divergence, work-groups should be organised so that warps follow the same control paths. In the case of Algorithm~\ref{alg:node_index} in the Appendix, thread divergence occurs when, as is most often the case, the number of non-broken bonds between nodes differs. The warp will not complete until all work-items in that warp have completed the \textbf{\texttt{for}} loop starting at line 5. This results in many threads in the warp being under-utilised during that time. However, in \texttt{PeriPy}, Algorithm~\ref{alg:bond_index} in the Appendix, the parallelisation is over bonds, and so the time that threads are underutilised is limited to the amount of time it takes to calculate a single bond rather than a number of bonds.

\smallskip
The problem of thread divergence has not been eliminated in Algorithm~\ref{alg:bond_index}. For example, if some work-items in warp execute the \texttt{if} statement in an \texttt{if-then-else} construct on lines 4-16 (calculating the bond force) while others follow the else path (the bond is broken), the GPU can no longer execute the work-items concurrently, and multiple passes are required. It is not possible to prevent thread divergence and to fuse the kernel that checks if a bond has exceeded the critical stretch into the kernel that calculates the bond force, as both the if and the else paths of the control loop will be executed by many work-items. Clearly, there is a trade off in the performance gain between kernel fusing and preventing thread divergence. Preliminary tests showed that kernel fusing at the cost of thread divergence in this manner is the faster approach across all problem sizes, by two orders of magnitude for large problem sizes $n>10^6$. This is because it avoids the memory constraints relating to storing bond forces in an $n \times N$ \texttt{C} array in global memory at every time step, the bandwidth constraints of accessing this array in a separate check-bonds kernel and the need to calculate the bond stretch, $s$, two times.

\smallskip
Increasing $n$ by a factor of 2 will double the amount of warps that need to be scheduled and executed and will thus decrease performance by a factor of 2. The amount of warps that need to be executed per node will be the dominant factor for per-node performance. Increasing $N$ by a factor of 2 will double the number of warps that need to be executed and thus decrease performance by approximately a factor of 2. This relationship is approximate since not all warps take equal time to execute; for example, warps that contain all broken or 'padding' bonds will take much less compute time to execute compared to warps with non-broken bonds.
% In other words, performance will scale linearly with $n$ and approximately linealy with $N$.

\subsection{Overview of Code Design}\label{design_overview}
\begin{figure}
	\centering    
	\includegraphics[width=9.0cm]{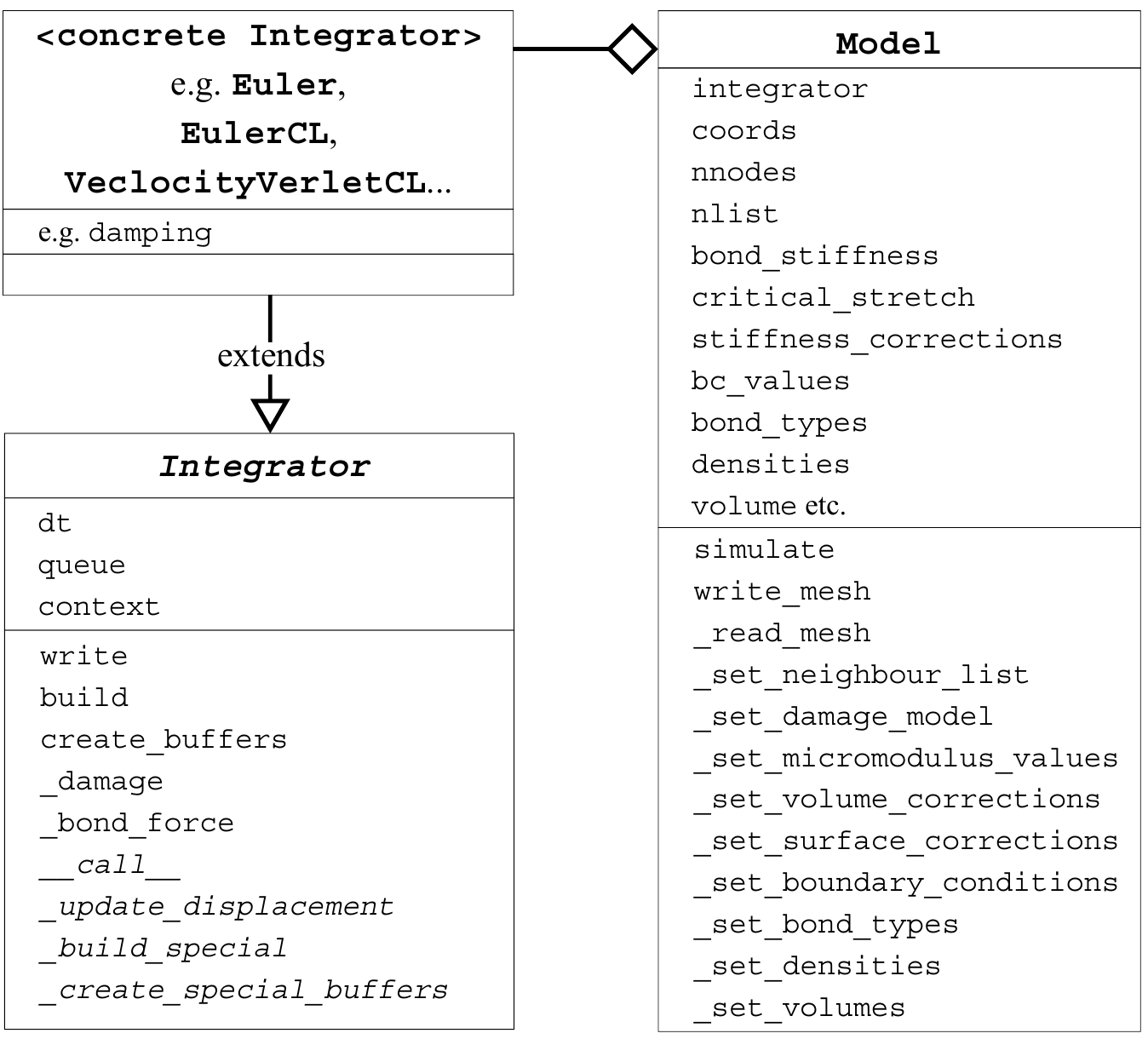}
	\caption{The class diagram for \texttt{PeriPy}, which is designed to easily extend the code with new integrators, correction algorithms or disretisations. The functionality of the code is defined in two classes: \texttt{Integrator}, which is the Abstract Base Class (ABC) that is extended by a concrete Integrator class, such as \texttt{EulerCL}, and \texttt{Model}, which defines the geometry (mesh and node volumes), connectivity (neighbour list and number of neighbours for each node), damage model, material types, correction algorithms and boundary conditions.}
	\label{fig:class_diagram}
\end{figure}

The class diagram for \texttt{PeriPy} is presented in Fig.~\ref{fig:class_diagram} and shows the \texttt{peripy.model.Model} class, which allows users to define a bond-based peridynamics model for composite materials with non-linear damage models defined by an arbitrary number of linear splines, such as the trilinear damage model illustrated in Fig.~\ref{fig:trilinear_damage_model}. Surface correction factors, partial volume correction factors, different micromodulus functions and boundary conditions can be readily included by supplying the relative keyword arguments to the model class. The model is defined by a damage model, the set of keyword arguments described above, a set of initial conditions (coordinates, connectivity and, optionally, node state vectors), force and/or displacement controlled boundary conditions and a set(s) of nodes the user has chosen to measure a datum for. The design of the code has facilitated the following features.\smallskip

 \noindent\textbf{Readily extended integrators:} When instantiating the \texttt{peripy.model.Model} class, an explicit integrator that inherits \texttt{peripy.integrators.Integrator} as an Abstract Base Class is required. \texttt{PeriPy} has not yet been extended to implicit integration methods, and contributions are welcome. Examples of the explicit integrators included are: \texttt{peripy.integrators.Euler}, \texttt{peripy.integrators.EulerCL}, \texttt{peripy.integrators.EulerCromerCL} and \texttt{peripy.integrators.VelocityVerletCL}. The suffix \texttt{CL} represents an integrator that utilises \texttt{OpenCL} kernels. As mentioned earlier, the \texttt{CL} integrators can be utilised on any CPU or GPU platform due to the heterogeneous nature of \texttt{OpenCL}.\smallskip

 \noindent\textbf{Regular and irregular meshes:} \texttt{PeriPy} uses the python package \texttt{meshio} \cite{meshio} to read input meshes of various formats, where the vertices of the meshes are the peridynamic nodes. \texttt{PeriPy} can take cuboidal tensor grid meshes as input, which are often employed so that the nodal volumes $dV_i$ can be easily and accurately calculated. Regular or irregular tetrahedral grids may also be used, with the only complication being the need to define a volume associated with each node. This may be accomplished by using computational geometry techniques, such as Delaunay triangulation or Voronoi diagrams \cite{Okabe2000}, which are not included in the package. Instead, if a tetrahedral mesh is used and exact nodal volumes are not supplied through the keyword argument \texttt{volume} of the model class \texttt{peripy.model.Model}, \texttt{Peripy} will approximate the nodal volumes by a weighted sum of the volumes of the tetrahedra of which the node is a vertex. Specifically, if a node is a vertex of a tetrahedra, it will be assigned 1/4 of the volume of that tetrahedra. The volumes of the tetrahedra are calculated using the standard formula of a dot product and cross product of the difference of the vertex position vectors. With this method, the total sum of the nodal volumes is equal to the total volume of the mesh, and as the volumes go to zero the absolute error is reduced. In the case of a regular cuboid grid that is split into tetrahedra, as the mesh elements are congruent, this method will result in exact volumes in the bulk. However, on the boundaries there will always be a significant error in the volumes when compared to the Veronoi diagram method. This results in surface nodes that have different effective stiffnesses to the bulk and will cause surface effects extraneous to the `peridynamics surface effect' explored by Le and Bobaru \cite{Le2018}. It is therefore recommended that, if an irregular mesh is used and the exact volumes are known, then they should be supplied through the keyword argument \texttt{volume}. While both irregular and regular tetrahedral meshes are possible with \texttt{PeriPy}, for the purposes of this paper, regular cuboidal tensor grid meshes were used, keeping with the majority of the existing literature.\smallskip

 \noindent\textbf{Composite materials:} Composite materials and structures can be defined and supplied to the \texttt{peripy.model.Model} class.\smallskip

 \noindent\textbf{General boundary conditions:} General boundary conditions can be supplied, and magnitudes of the boundary conditions can be changed over time. The directions of the boundary conditions cannot be changed over time, except along the directions that they were initially defined. This means that, while most commonly used boundary conditions are possible, torsion, for example, is not currently possible; however, contributions are welcome.\smallskip

 \noindent\textbf{Reduced overhead:} The neighbour list generated using the \texttt{scipy.spatial.KDTree} class \cite{2020SciPy-NMeth}. In \texttt{PeriPy}, the arrays that are constructed during the instantiation of the \texttt{peripy.model.Model}, such as the neighbour list, can be written to an h5 \cite{HDF5} file and read by the \texttt{peripy.model.Model} through keyword arguments. This reduces the overhead needed to simulate an example.

 \noindent\textbf{Running simulations:} Simulations are conducted with \texttt{Model.simulate}. The peridynamics parameters of a damage model, node states, initial conditions and boundary conditions can be changed without reinstantiating the \texttt{peripy.model.Model} by setting them as keyword arguments to the \texttt{peripy.Model.simulate} method. In this way, multiple simulations with different parameters be conducted with minimal overhead cost so that 'outer loop' applications, including sensitivity analysis, uncertainty quantification and optimisation, may be performed.\smallskip

 \noindent\textbf{Initiating a simulation from an intermediate state:} The \texttt{peripy.model.simulate} can be restarted from an intermediate state by supplying a \texttt{first\_step} argument and state variables for that time step. This could be useful, for example, when changing parameters that do not affect the elastic response of the model but do affect the crack nucleation and propagation, in which case the simulation states would be saved in the linear elastic region of a quasi-static response of an increasing load.

 \noindent\textbf{Output visualisation in ParaView:} The simulate method will write the states to a mesh file that is easily visualised (e.g. Paraview \cite{paraview}). Writes occur at a specified frequency, in a number of steps. The other outputs of the method are a tuple of the state vectors and a dictionary object containing the average of displacement, velocity and acceleration as well as the sum of forces and body forces for each of the writes (over time), for each unique \texttt{tip\_type}, which is defined as the set of nodes the user has chosen to measure a datum for, as defined by the \texttt{is\_tip} function. This could be the set of nodes on the tip or midspan of a structure.\smallskip

For a comprehensive explanation and usage of all classes, methods and arguments, please see the latest full documentation at \href{https://peripy.readthedocs.io/en/latest/}{\texttt{https://peripy.readthedocs.org}}.

\section{Benchmarking}\label{benchmarking}
This section will compare the performance of the solver versus two existing solvers. One is an \texttt{OpenMP} implementation by \citet{Hobbs2020}. All the functions are written in \texttt{C} and use \texttt{OpenMP}. The \texttt{C} functions are called directly by \texttt{MATLAB}. This requires compilation of the \texttt{C} code into a \texttt{MEX} function (\texttt{MATLAB} executable). The compilation used \texttt{gcc-6.3.0} for Linux and the \texttt{MATLAB} version used was \texttt{R2018b}. The computations were performed using a single Skylake node with 32 OMP threads with the Cambridge Service for Data Driven Discovery (CSD3) Peta4 platform operated by the University of Cambridge Research Computing Service. The other solver is the \texttt{OpenCL} implementation by \citet{Mossaiby2017}. The source code provided by \citet{Mossaiby2017} was executed by us along with the \texttt{OpenMP} and \texttt{PeriPy} codes. Both \texttt{OpenCL} solver computations were run on an NVIDIA 2080Ti with 11GB GDDR6 RAM and 4352 CUDA cores, and \texttt{pyopencl==2020.1} was used.

\smallskip
The results show a 1.4-2.0x performance gain over the \texttt{OpenCL} solver, with an up to 10.0x performance increase for problems with a small number of nodes $n$. The results also show a 3.7-7.3x performance gain over the \texttt{OpenMP} solver. Linear compute scaling across problem size $n$ (number of nodes) and family size $N$ is demonstrated. There is a negligible difference in the performance of \texttt{PeriPy} as a result of using stiffness correction factors.% The performance of double versus single precision is shown.
The performance difference of using an irregular mesh rather than a regular mesh is shown and an explanation provided.

\subsection{Comparing across problem size, and horizon distance}
The performance of \texttt{PeriPy} across problem size with and without any stiffness correction factors is shown in Fig.\ref{fig:profile_regular}. The performance depends on the number of warps that need to be executed on the device. The performance is linear with the number of nodes $n$ and approximately linear with the family size $N$, for the reasons discussed in Section~\ref{thread_divergence}.

\begin{figure}
	\begin{subfigure}{0.5\textwidth}
		\includegraphics[width=9.0cm]{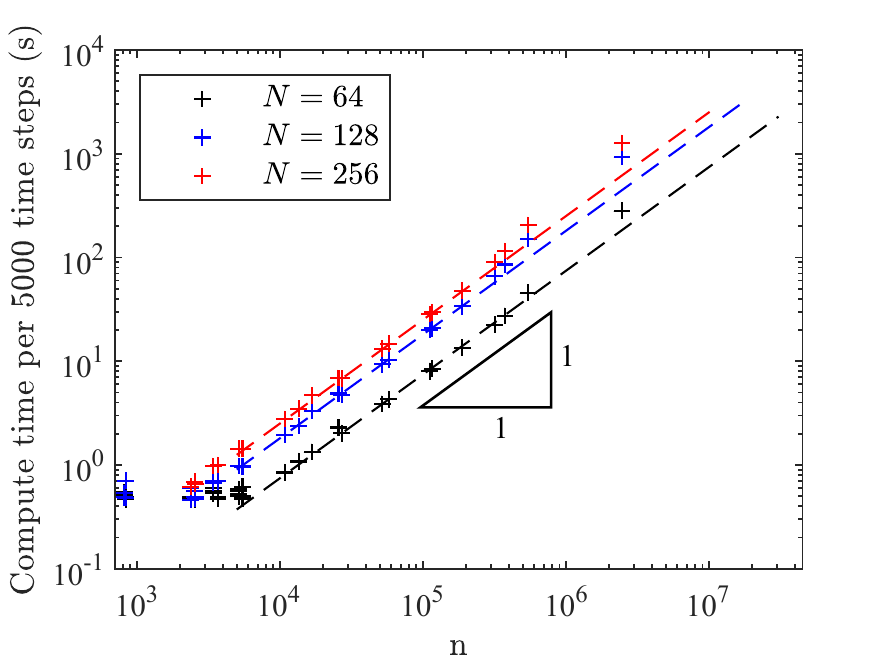}
		\caption{Without stiffness correction factors.}
		\label{fig:profile_no_irregular}
	\end{subfigure}
	\begin{subfigure}{0.49\textwidth}
		\includegraphics[width=9.0cm]{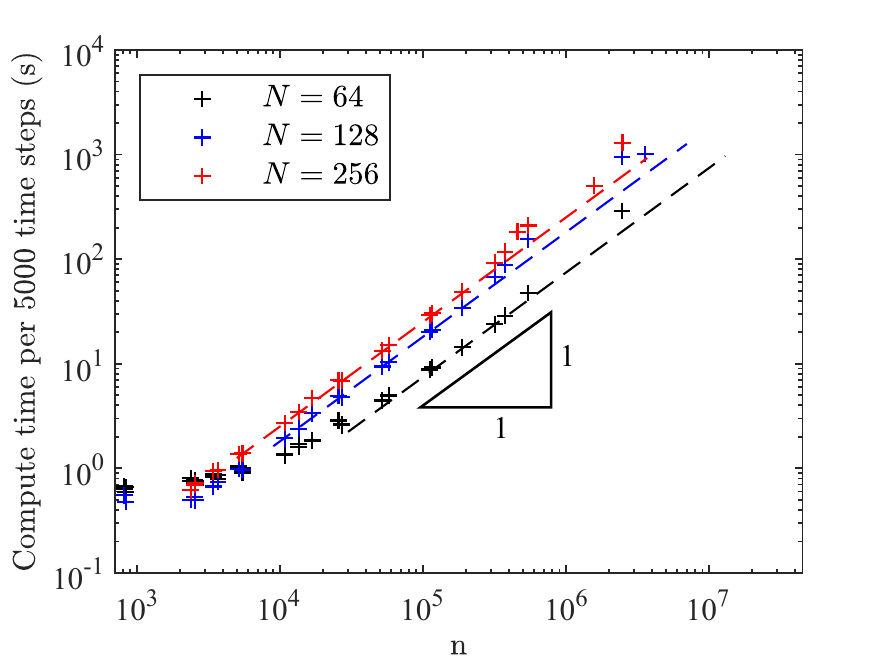}
		\caption{With stiffness correction factors}
		\label{fig:profile_s_irregular}
	\end{subfigure}
	\hfill
	\caption{\texttt{PeriPy} performance across problem size for 5000 time steps with an irregular mesh. $n$ is the number of nodes and $N \geq \max_{i}(|\mathcal{H}_i|)$ is the maximum number of nodes in the non-local family. A number of problem sizes were used to demonstrate the linear scaling performance of \texttt{PeriPy} for problems with a large number of nodes. The benchmark tests were run on an NVIDIA 2080Ti GPU. The particular geometries and boundary conditions do not affect performance, and simple ones were applied.}
	\label{fig:profile_irregular}
\end{figure}

\smallskip
The performance with irregular meshes is shown in Fig.\ref{fig:profile_irregular}. For the same $N$, performance for the irregular meshes is faster than with regular meshes. This is because, in irregular meshes, the range of the family size $|\mathcal{H}_i|$ across the node index $i$ is larger than in a regular mesh, so the \texttt{group\_size}, $N = \min_{k}(2^k) \geq \max_{i}(|\mathcal{H}_i|)$, is typically 2 times larger than the regular mesh with equivalent average spacing $\Delta x$. As a result, there are whole warps in the irregular mesh case for which no bond force calculations are required, which are performed much faster time than warps for which a bond force calculation is required. As a result, the performance for an $N = 64$ irregular mesh closely matches that of $N=32$ with a regular mesh; the performance for an $N = 128$ irregular mesh closely matches that of $N=64$ with a regular mesh and so on, as shown in Table~\ref{table:profile_gradients}.
\begin{table}[H]
	\caption{The number of additional nodes to increase the wall clock time of 5000 time steps by one second, $\frac{\partial n}{\partial t}\bigg|_{N}$.}
	\centering
	\label{table:profile_gradients}
	\begin{tabular}{l c c}
		\toprule
		 N & Regular mesh & Irregular mesh \\ 
		\midrule
		$32$ & $11487$ & \\
		
    	$64$ & $6452$ & $13341$\\
	
		$128$ & $3652$ & $5532$\\
		
		$256$ & $2500$ & $3990$\\
		\bottomrule
	\end{tabular}
\end{table}

\smallskip
The effect on performance of using stiffness correction factors (i.e. through micromodulus function, partial volume correction or surface corrections) is negligible, except at small problem sizes. At smaller problem sizes, the overhead due to memory transfer between the GPU and controlling CPU dominates, and copying the stiffness correction factor \texttt{C} array buffer onto GPU impacts performance. At larger problem sizes, the time required for actual computation comprises most of the total wallclock time, and the only change in the computation is the indexing of a constant stiffness correction factor array, which does not significantly impact performance.

\subsection{Comparing against existing solvers}

\begin{figure}[H]
	\begin{subfigure}{0.5\textwidth}
		\includegraphics[width=9.0cm]{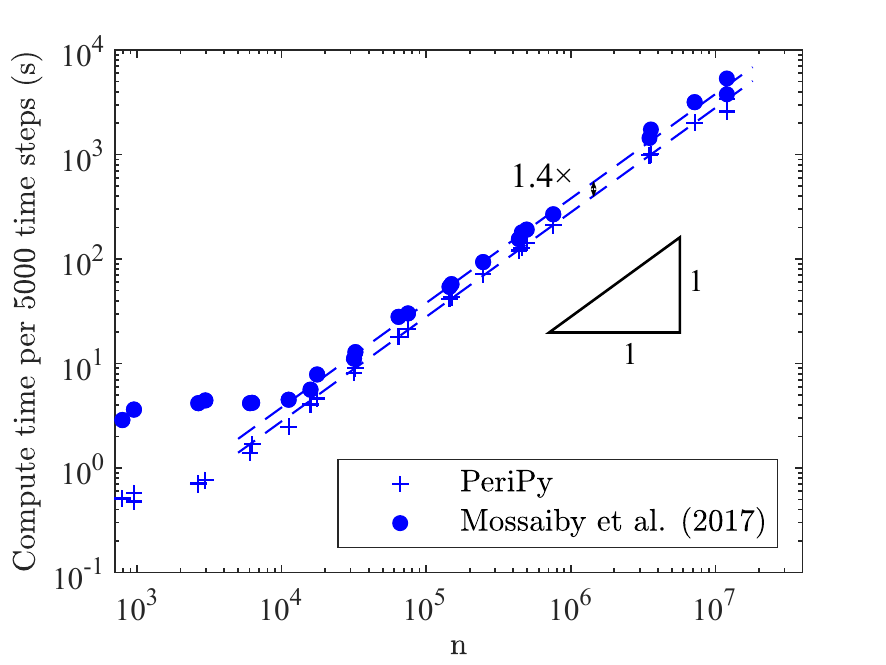}
		\caption{$N=128$, regular mesh}
		\label{fig:128_regular}
	\end{subfigure}
	\begin{subfigure}{0.49\textwidth}
		\includegraphics[width=9.0cm]{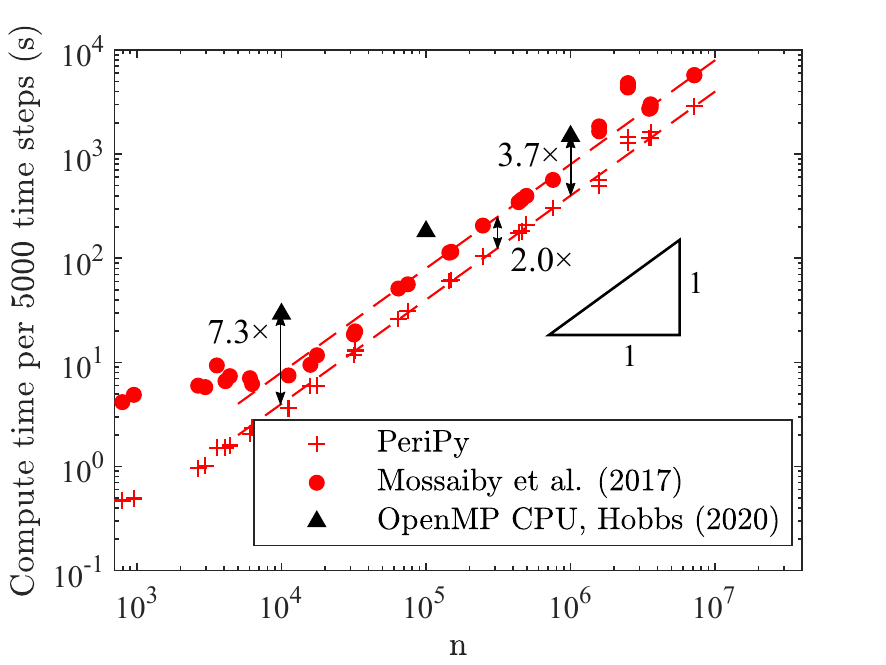}
		\caption{$N=256$, regular mesh}
		\label{fig:256_regular}
	\end{subfigure}
	\hfill
	\caption{Comparison across different implementations for 5000 time-steps with a regular mesh.}
	\label{fig:implemenations_regular}
\end{figure}

The performance of \texttt{PeriPy} when compared to the \texttt{OpenCL} solver by \citet{Mossaiby2017} is shown for the \texttt{group\_size} $N=128$ in Fig.~\ref{fig:128_regular}, and with the added comparison against the \texttt{OpenMP} solver by Hobbs it is shown for the \texttt{group\_size} $N=256$ in Fig.~\ref{fig:256_regular}. The speedups identified in Section~\ref{implementation} result in \texttt{PeriPy} being 1.4-2.0 times faster than the \texttt{OpenCL} solver by \citet{Mossaiby2017} and 3.7-7.3 times fast than the \texttt{OpenMP} solver by Hobbs. As explained in Section~\ref{thread_divergence}, the speed increases are a result of (1) kernel fusing of the \texttt{calc\_bond\_force} and \texttt{check\_bonds} kernels, (2) parallelisation over bonds (as opposed to nodes), which facilitates the most significant performance enhancement which is (3) minimisation of thread divergence. The relative effect on the performance of the speedups and a detailed explanation are provided below.

\begin{figure}[H]
	\begin{subfigure}{0.5\textwidth}
		\includegraphics[width=9.0cm]{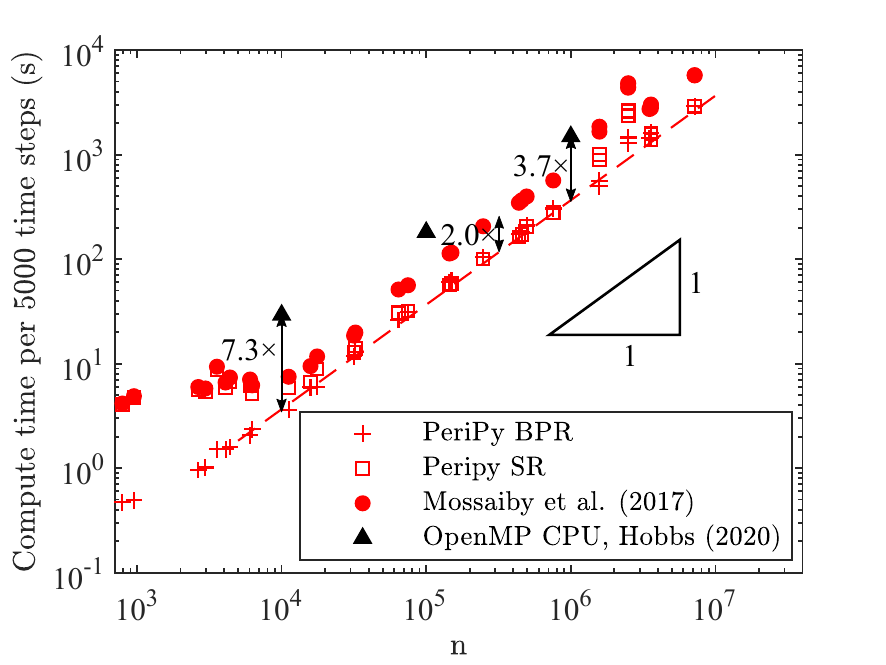}
		\caption{$N=256$, regular mesh}
		\label{fig:256_regular2}
	\end{subfigure}
	\begin{subfigure}{0.49\textwidth}
		\includegraphics[width=9.0cm]{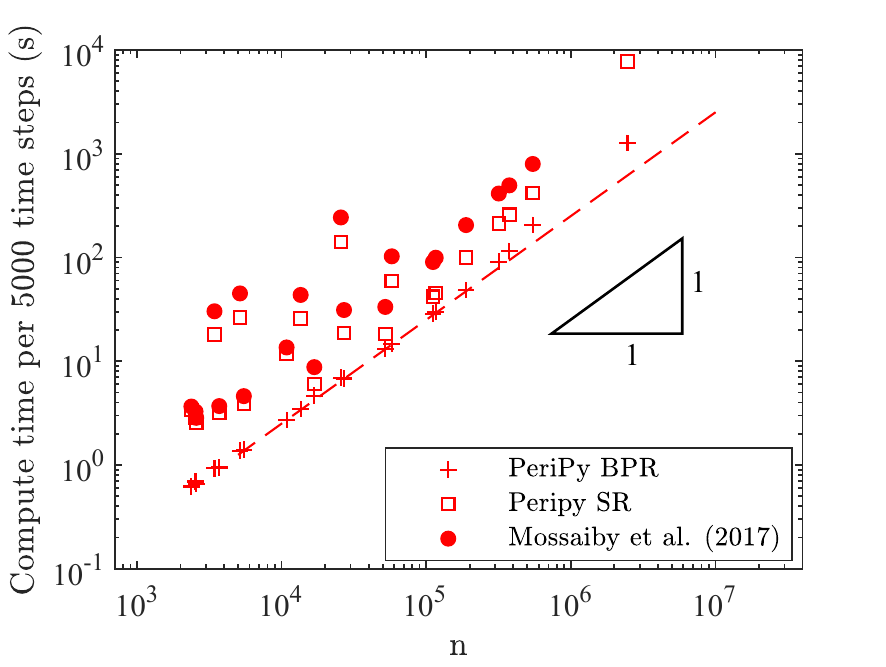}
		\caption{$N=256$, irregular mesh}
		\label{fig:256_irregular2}
	\end{subfigure}
	\hfill
	\caption{Irregular mesh versus regular mesh comparison across different implementations for 5000 time-steps.}
	\label{fig:implementations_256}
\end{figure}

\smallskip
Fig.~\ref{fig:implementations_256} shows the effect of binary parallel reduction (\texttt{PeriPy BPR}) on performance when compared to a serial reduction (\texttt{PeriPy SR}). Mossaiby et al. (2017) do not include any of the optimisations, while \texttt{PeriPy SR} has optimisations (1) but not (2) or (3). The speedup due to (1) is significant and is shown by the difference between \texttt{PeriPy SR} and Mossaiby et al.

\smallskip
The logical implications of the results in Fig.~\ref{fig:256_regular2} are that the summation algorithm, parallel or serial, does not actually yield much performance gain in and of itself. Rather, the binary reduction facilitates the parallelisation over bonds without needing to store a huge bond forces array in the GPU global memory every time step in order to do a summation of bond forces. The majority of the time savings for regular meshes is a result of kernel fusing.

\smallskip
However, it is for the irregular mesh, as shown in Fig.~\ref{fig:256_irregular2}, that the speedups due to (2) and (3), which are implemented in \texttt{PeriPy BPR}, are fully realised. These speedups are primarily due to reducing thread divergence to a practical minimum by parallelising over bonds rather than nodes. Consider, as is the case in \texttt{PeriPy SR}, parallelising over nodes. Each work-item, corresponding to a node, has to calculate the force of non-broken bonds for that node. It is often the case for irregular meshes that the number of non-broken bonds between each node varies significantly. Therefore, parallelising over nodes results in significant thread divergence, which depends on the range of broken bonds for the nodes. In \texttt{PeriPy BPR}, we reduce the thread divergence as much as possible by parallelising over bonds. The only thread divergence left is whether or not the bond force for a bond is calculated, and hence we achieve the minimum possible thread divergence for the problem, which results in the speedup. This explains why the performance of \texttt{PeriPy} for irregular meshes is much better than in Mossaiby and the \texttt{PeriPy SR} implementation.

\smallskip
It can also be seen from Fig.~\ref{fig:implementations_256} that, for small problems, where the  overhead due to memory transfer between the GPU and controlling CPU dominates, \texttt{PeriPy BPR} is up to 10 times faster than Mossaiby et al. and \texttt{PeriPy SR}. The difference between \texttt{PeriPy SR} and \texttt{PeriPy BPR} is that the parallelisation is over bonds rather than nodes (2), and so forces are calculated in local rather than global memory, reducing communication time, and the thread divergence is also reduced (3).

%\subsection{Comparing single versus double precision performance in \texttt{PeriPy}}
%
%\begin{figure}
%	\begin{subfigure}{0.49\textwidth}
%		\includegraphics[width=9.0cm]{precision_regular_128.pdf}
%		\caption{$N=128$, linear performance with increasing problem size. Double precision is approximately 8.6 times slower than single precision.}
%		\label{fig:precision_128}
%	\end{subfigure}
%	\begin{subfigure}{0.49\textwidth}
%		\includegraphics[width=9.0cm]{precision_regular_256.pdf}
%		\caption{$N=256$, linear performance with increasing problem size. Double precision is approximately 6.7 times slower than single precision.}
%		\label{fig:precision_256}
%	\end{subfigure}
%	\hfill
%	\caption{Comparison single precision vs double precision.}
%	\label{fig:precision}
%\end{figure}
%
%Fig.~\ref{fig:precision_128} shows a comparison of the performance of \texttt{PeriPy} for double and single precision. The gradients of the linear region are shown in Table~\ref{table:profile_gradients_precision}.
%\begin{table}[H]
%	\caption{The number of additional nodes to increase the wall clock time of 5000 time-steps by one second, $\frac{\partial n}{\partial t}\bigg|_{N}$.}
%	\centering
%	\label{table:profile_gradients_precision}
%	\begin{tabular}{l c c}
%		\toprule
%		 N & Single precision & Double precision \\ 
%		\midrule
%		$128$ & $32258$ & $3916$\\
%		$256$ & $16667$ & $2500$\\
%		\bottomrule
%	\end{tabular}
%\end{table}

\section{Validation of \texttt{PeriPy} through engineering test cases}\label{validation}
In this section we demonstrate the use of \texttt{PeriPy} for explicit schemes for a 3D dynamic test. In testing, \texttt{PeriPy} was also validated against other solvers for simple test cases.

\subsection{Example 1: Kalthoff--Winkler experiment}\label{Kalthoff-Winkler}
\begin{figure}
	\centering    
	\includegraphics[width=0.6\textwidth]{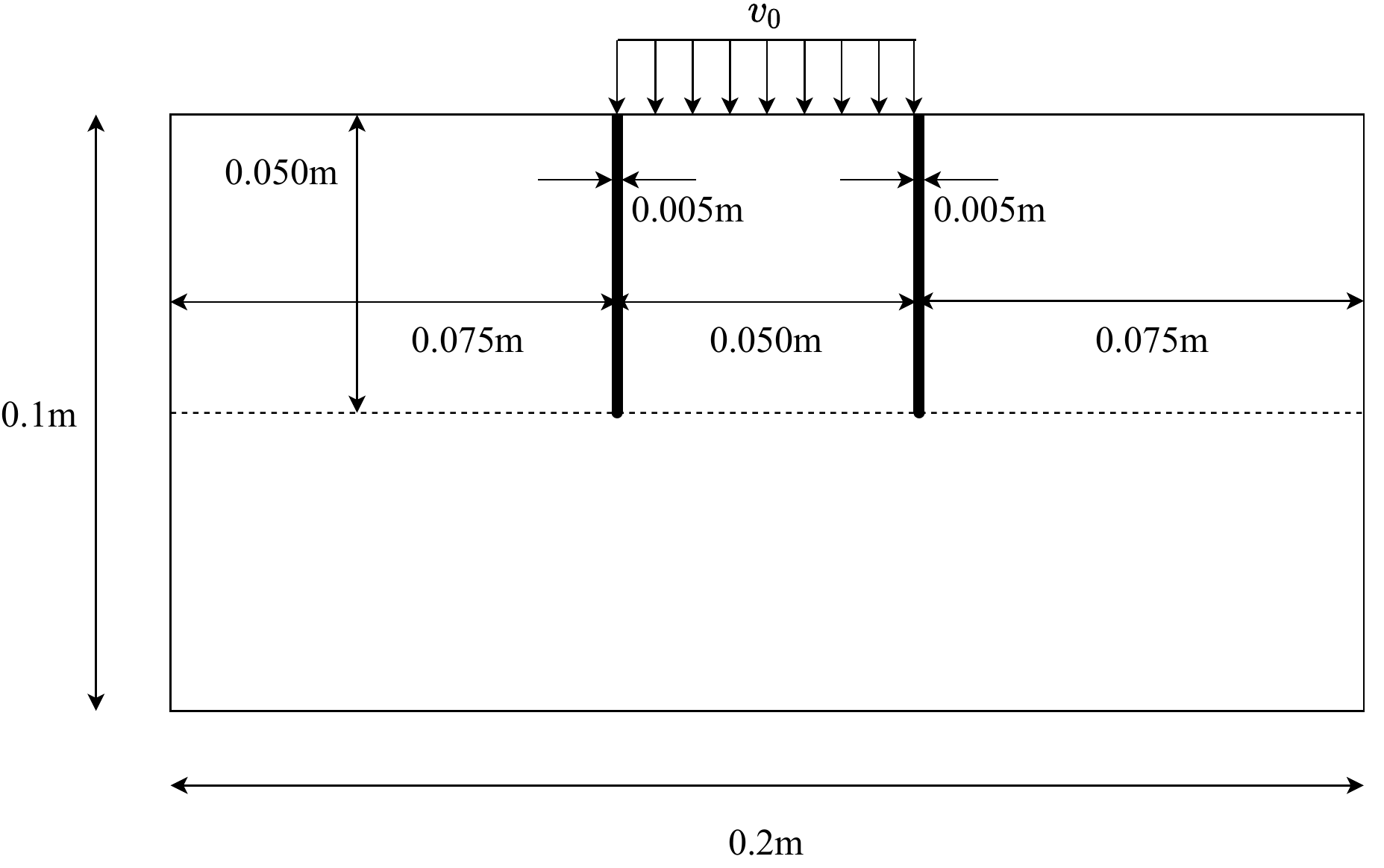}
	\caption[Crack speed]{Setup of Kalthoff--Winkler's experiment.}
	\label{fig:Katlhoff-Winkler}
\end{figure}

The Kalthoff--Winkler experiment \cite{Kalthoff1988} \cite{Kalthoff2000} is a classical benchmark problem for dynamic fracture \cite{Belytschko2003, Rabczuk2010, Song2006}. The geometry and boundary conditions used for the simulation are illustrated in Fig.~\ref{fig:Katlhoff-Winkler}. The thickness of the specimen is \SI{9}{\milli \metre}. For a plate made of 18Ni1900 steel subject to impact loading at the speed of $v_0 = \SI{32}{\metre \per \second}$, brittle fracture was observed with a crack forming at \ang{68} angles to the notches. We repeat the simulation first conducted by Silling \cite{Silling2003} and repeated by \citet{Ren2016} using the same material parameters as in the latter, comparing our results for validation purposes. The elastic modulus is $E = \SI{190}{\giga \pascal}$, the density is $\rho = \SI{7800 }{\kilo \gram \per \metre \cubed}$, Poisson's ratio is $\nu = 0.25$ and the energy release rate is $G_0 = \SI{6.9e4}{\joule \per \metre \squared}$. Repeating the experiment by \cite{Ren2016}, impact loading was imposed by applying an initial velocity at $v_0 = \SI{22}{\metre \per \second}$ to the first three layers of nodes in the domain, as shown in Fig.~\ref{fig:Katlhoff-Winkler}; all other boundaries are free. The plate is discretised with a node diameter $\Delta x = \SI{1.5625}{\milli \metre}$. Along the thickness of the plate, four layers of nodes are employed to match the simulation in \cite{Ren2016}. The total number of nodes is $n=32,768$.

\smallskip
We also repeat the simulation conducted by \citet{Mossaiby2017} in the validation of their solver, using material parameters the same as in the study by \citet{Ren2016}, but, following \citet{Mossaiby2017}, the plate is discretised with a node diameter of $\Delta x = \SI{1./2560}{\metre}$, with $13$ layers of nodes, contributing to the total number of nodes $n=1,713,933$.

\begin{figure}
	\centering    
	\includegraphics[width=9.0cm]{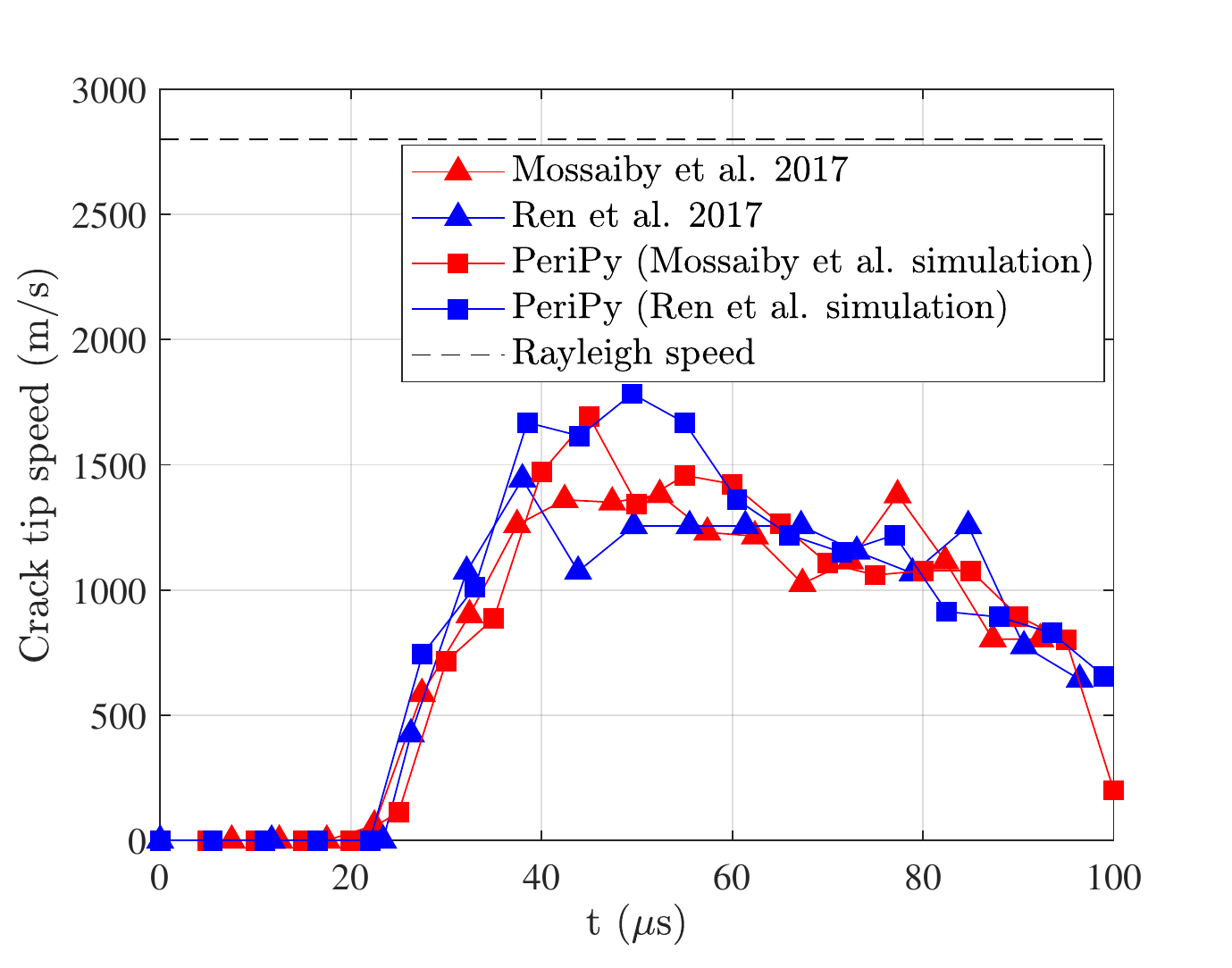}\includegraphics[width=9.0cm]{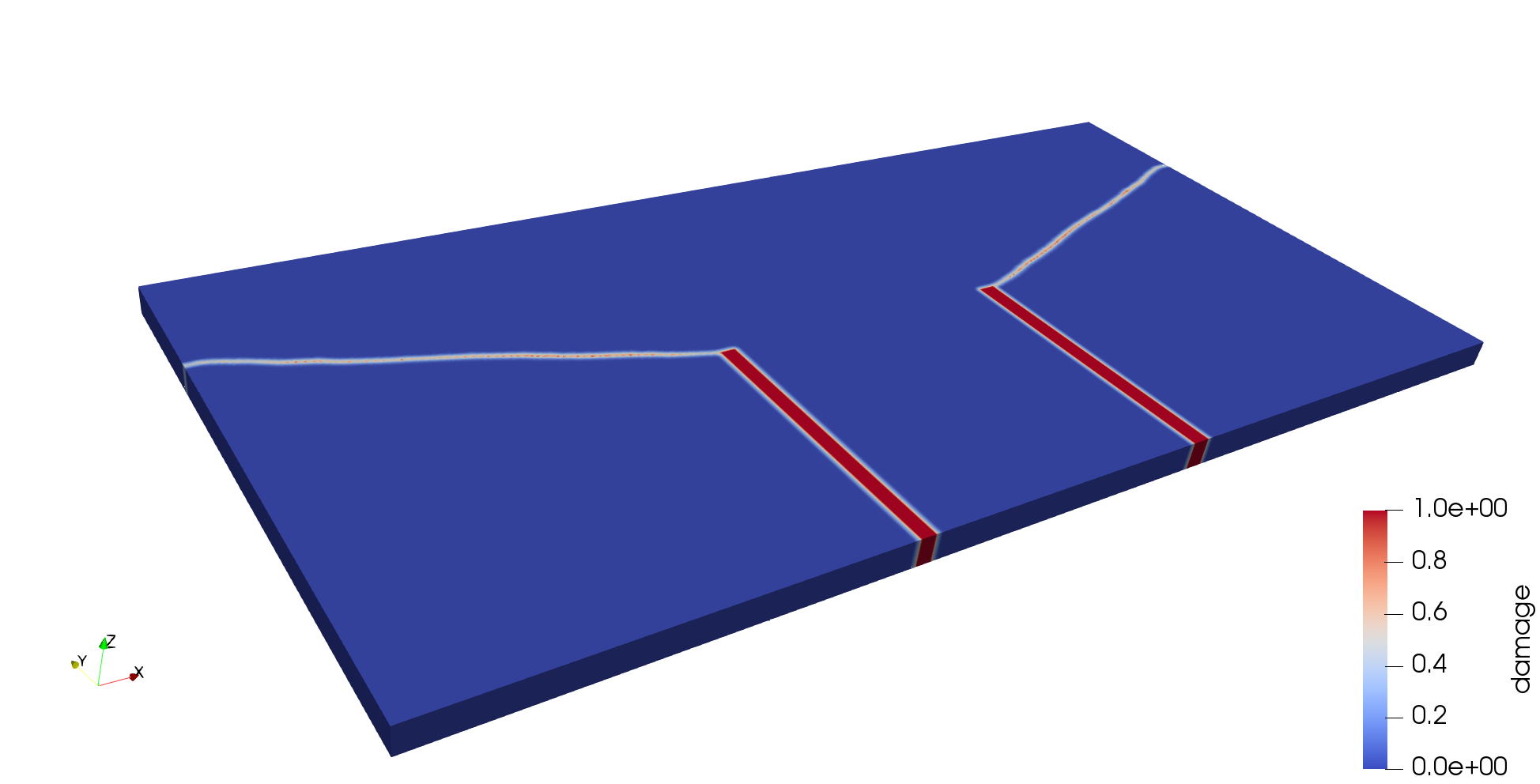}
	\caption[Crack speed]{(Left) The crack propagation speed in Example 2. (Right) The crack at \SI{100}{\micro\second} as viewed as a mesh file in ParaView \cite{paraview}.}
	\label{fig:crack_speed_across_simulations}
\end{figure}

\smallskip
It should be noted here that, in order to prevent crack nucleation at the centre of the long edge away from the pre-cracks, a `no failure zone' must be imposed on that edge, as can be seen in the simulation of \cite{Ren2016}. If no `no failure zone' is to be used, then a micro-elastic plastic damage model must be used, as mentioned by Silling \cite{Silling2003}. The `no failure zone' is simple to implement in \texttt{PeriPy} and results in a fracture pattern that matches the simulations by \citet{Ren2016} and \citet{Mossaiby2017}, the latter of which is shown in Fig.~\ref{fig:crack_speed_across_simulations} (right).\smallskip

The crack propagation speed is given as
\begin{equation}
\|\dot{u}(t + \frac{\Delta t}{2})\| = \frac{\|\bm{x}_{\text{tip}}(t + \Delta t) - \bm{x}_{\text{tip}}(t)\|}{\Delta t}
\end{equation}
where $\bm{x}_{\text{tip}}(t)$ is the position of the crack tip at time $t$. The Rayleigh wave speed $c_R$ \cite{Graff1975} is given as
\begin{equation}
\frac{c_R}{c_s} = \frac{0.87 + 1.12 \nu}{1 + \nu}
\end{equation}
where $c_s = \sqrt{\mu / \rho}$ is the shear wave speed, and $\mu$ is the shear modulus.\smallskip

The crack tip is measured using the corner detection algorithm developed by Harris\cite{Harris1988} with the OpenCV python package \cite{opencv_library}. Corner detectors can be unreliable, and higher-level visual routines must be designed to tolerate a significant number of
outliers in the output of the corner detector. Outliers were filtered if the resulting velocity was above the Rayleigh speed, and the results were averaged with slightly different values for the free parameter $\kappa$, which describes how `edge phobic' the corner detection algorithm is. Fig.~\ref{fig:crack_speed_across_simulations} shows the evolution of the crack speed using this method. The speed was calculated every $\SI{5e-6}{\second}$. For verification purposes, our results are compared with those of the peridynamic model in \cite{Ren2016}.\smallskip

The crack starts to propagate at $\SI{24e-6}{\second}$, the same time as in the simulation of \citet{Ren2016} The highest crack speed in the simulation measured by the corner detection algorithm was $\SI{1830}{\metre \per \second}$, $65.4\%$ of the Rayleigh speed, which was larger than Ren's value of $\SI{1500}{\metre \per \second}$. Most other velocities are in good agreement. The crack angle at the kink was measured to be \ang{65.3}, closely matching Ren's reported value of \ang{65.7}.\smallskip

\section{Example 2: Outer-loop calibration of peridynamics model parameters using experimental force-deflection data}\label{concrete_beams}

\begin{figure}[ht]
	\centering
	\includegraphics[width=0.5\linewidth]{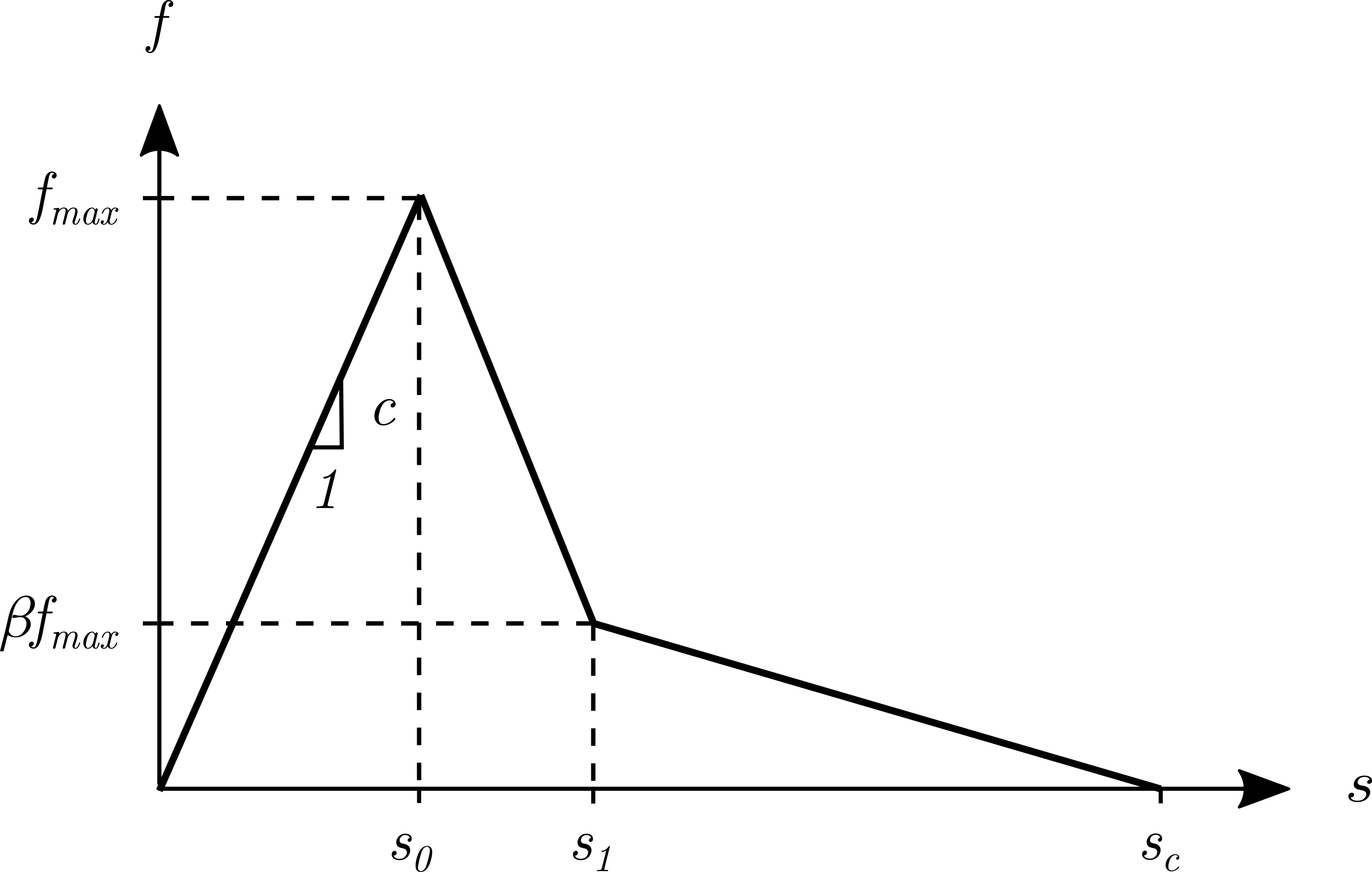}
	\caption{Trilinear damage model.}
	\label{fig:trilinear_damage_model}
\end{figure}

This section presents an optimisation problem to calibrate a bond-based trilinear damage model for a concrete beam simulation. Concrete is a quasi-brittle material and exhibits strain-softening behaviour when in tension. \citet{Yang2018} noted that the tension softening part of the bilinear model, was not based on the true softening behaviour of concrete and thus established a trilinear model within the bond-based peridynamic formulation, illustrated in Fig.~\ref{fig:trilinear_damage_model}. The shape of the softening branch in the trilinear model is consistent with the tension softening behaviour of concrete. A generalised method for determining the parameters $s_0$, $s_1$ and $s_c$ was proposed and only requires the elastic modulus, tensile strength and fracture energy. The position of the kink point in the softening curve is determined by the initial fracture energy, as proposed by Ba\v{z}ant \cite{Bazant2001}. However, the critical stretch of a peridynamic bond is dependent on fracture energy $G_F$. The selection of $G_F$ is a major source of uncertainty in peridynamic models. Furthermore, concrete material properties are typically determined through the testing of concrete compressive strength $f_c$. Additional material parameters, such as elastic modulus $E$, tensile strength $f_t$ and fracture energy $G_F$, are rarely tested and must be determined from empirical formulas. At present, there is no standard theory relating concrete compressive strength to fracture energy.

\begin{figure}[ht]
    \centering
    \includegraphics[width = 0.5\textwidth]{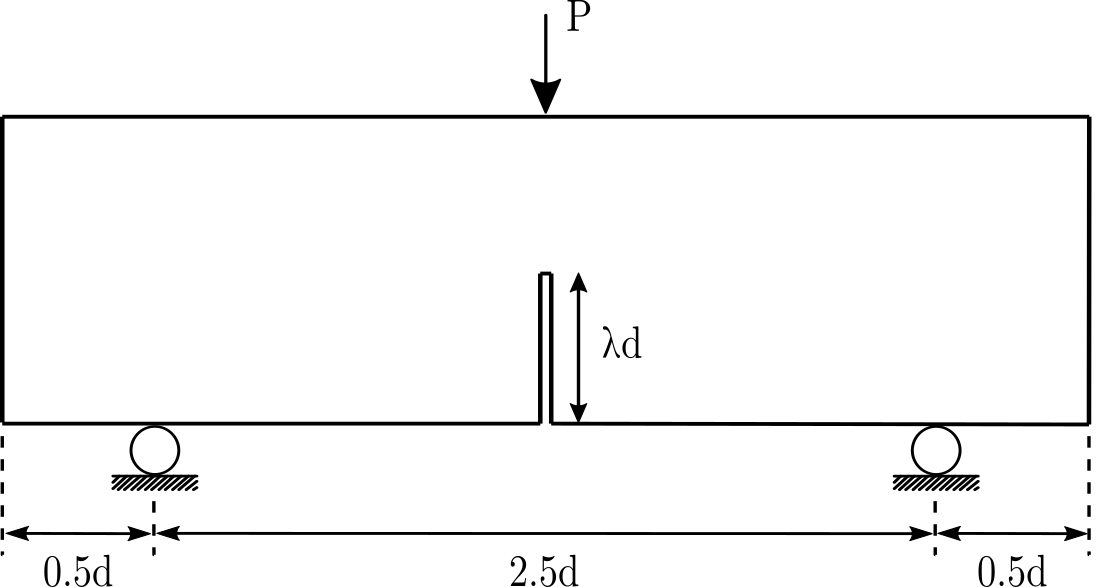}
    \includegraphics[width= 0.42\linewidth]{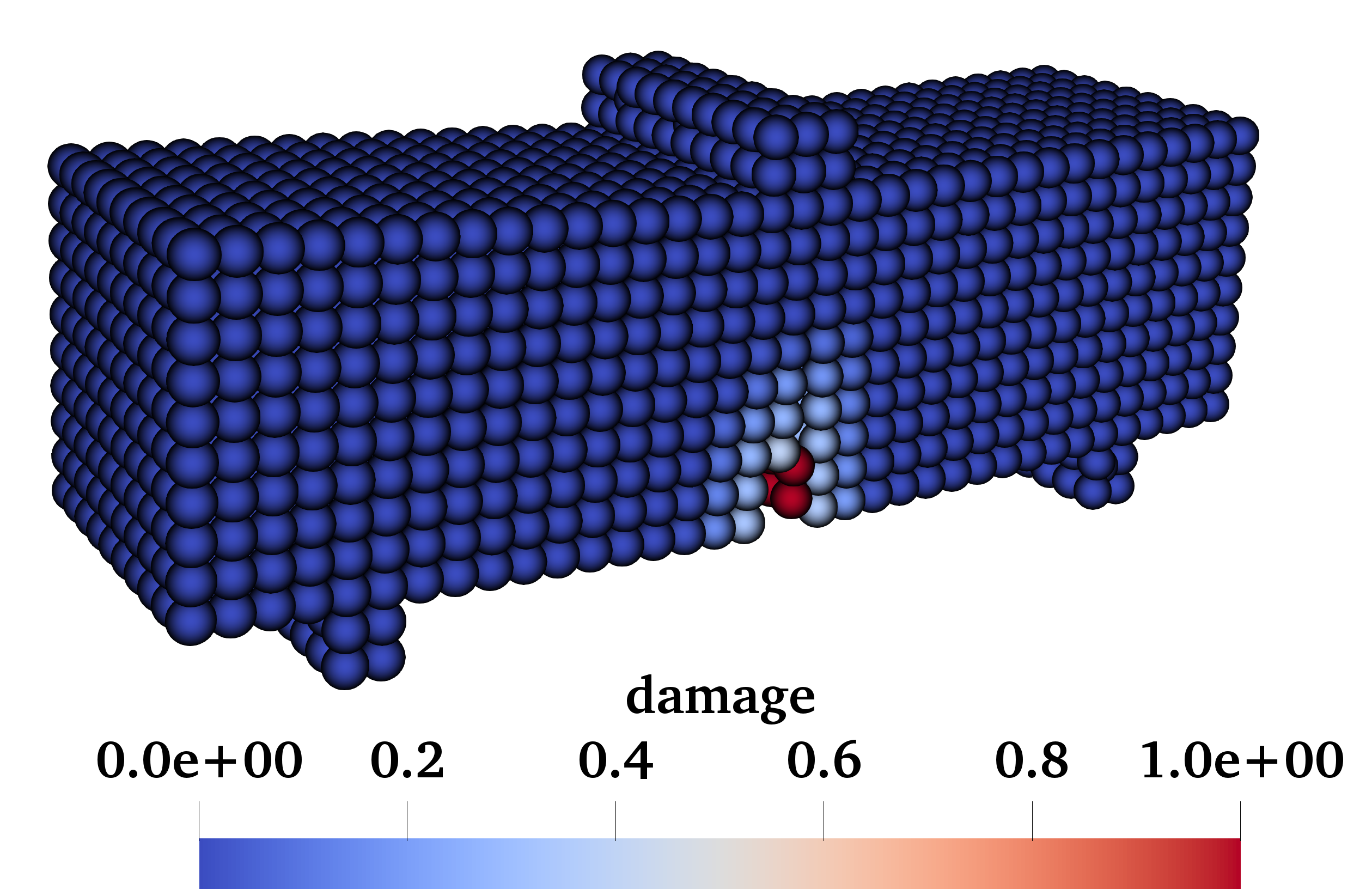}
    \caption{(Left) Schematic of the experimental setup reported by \citet{Gregoire2013}. (Right) Peridynamic model of fifth notched plain concrete beams, discretised with a $\SI{5}{\milli \metre}$ mesh. The boundary conditions are applied to the peridynamic nodes outside the surface of the beam, and the initial notch is shown in red.}
    \label{fig:Gregoire_experimental_setup}
\end{figure}

\begin{figure}[H]
	\includegraphics[width=9.0cm]{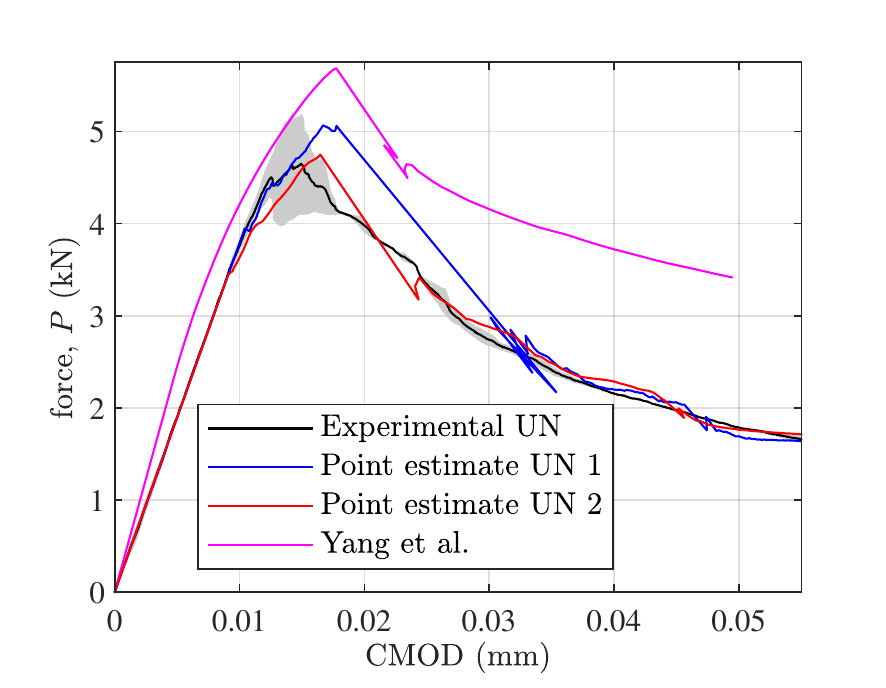}\includegraphics[width=9.0cm]{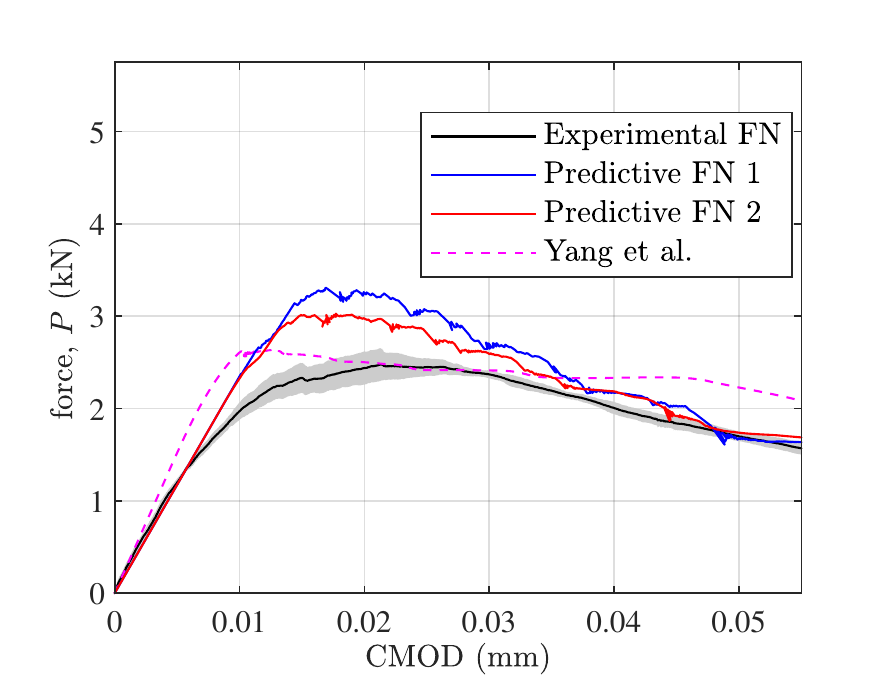}	\caption{For a $\Delta x = \SI{5}{\milli \metre}$ mesh, (a) experimental and calibrated curves, achieved by minimising the mean squared error between the experimental displacement plots and two point estimates of the trilinear parameters (coloured) on the un-notched (UN) beam geometry; (b) experimental and predictive curves, achieved by using the calibrated point estimates from the unnotched beam data applied fifth notched (FN) beam, i.e. $\lambda = 0.2$}
	\label{fig:calibration}
\end{figure}

\smallskip
It is therefore likely that an 'outer-loop' data-driven method, is required to calibrate the trilinear model against experimental data. \texttt{PeriPy} is a useful tool in such outer-loop applications, providing sufficiently fast model evaluates to enable optimisation routines. Following \citet{Yang2018}, we set the kink point to $\beta= 1/4$, and compare the values of the trilinear model of \citet{Yang2018} to optimised parameters that are calibrated against experimental data. A schematic diagram of the experimental setup used \cite{Gregoire2013} is illustrated in Fig.~\ref{fig:Gregoire_experimental_setup} (Left). Results for the unnotched (UN), (notch-to-depth ratio $\lambda$ = 0) and fifth-notched (FN), notch-to-depth ratio $\lambda = 0.2$ case are presented. The member has the following dimensions: length $l = 175$ mm; depth $d = 50$ mm; and thickness $b = 50$ mm. The span of the member is $125$ mm, and width of the notch is $2$ mm. The beam was tested multiple times, and the range of experimental results are plotted in grey in Fig.~\ref{fig:calibration} for the un-notched case and notched case. The means are plotted in black. The experimental tests are \emph{crack-mouth-opening-displacement} (CMOD) controlled to avoid unstable crack propagation after the peak load. The \textit{unnotched} plain concrete beam in three-point bending was used for calibration of point-estimates of the trilinear model parameters. The calibrated point-estimates were then tested on a \textit{notched} concrete beam and compared to experimental data.

The peridynamic model was  discretised using a $\Delta x = \SI{5}{\milli \metre}$ mesh giving a total number of peridynamic nodes of $n = 3,645$. A fixed peridynamic horizon of $\delta = \SI{15.7}{\milli \metre}$, and no stiffness correction methods were applied in this simulation (see Example 3). A displacement-controlled loading scheme is used for the numerical simulations. The left-hand support was clamped in all three directions. Following the experimental setup closely, the beam was simply supported. The right-hand support was constrained not to move in the vertical and out-of-plane directions. The central support was constrained not to move in the out-of-plane direction and had a prescribed displacement in the vertical direction. As noted by \citet{Mehrmashhadi2019}, applying imposed displacements to a single node will result in a significant peridynamic surface effect. Boundary displacements were applied over a two-by-two node cross-section outside the surface of the beam (except for the central point load, where the boundary was applied over a three-by-two node cross-section, for symmetry), which is shown in Fig.~\ref{fig:Gregoire_experimental_setup} (right). The boundary nodes were set as `no-failure-zones' so that they do not exceed critical stretch, which would be nonphysical, and their bond stiffness was taken to be the same as the bulk material. The displacement boundary conditions were applied smoothly using a fifth-order polynomial displacement time curve, such that the boundary nodes had an acceleration of zero at the first time step and during crack nucleation and growth. Dynamic damping was used to dampen oscillations in the beam. These steps serve to increase the stability of simulation and ensure that crack growth occurs under quasi-static loading conditions. The exact dynamic parameters used for the simulations and the velocity of the applied displacement in the vertical direction are shown in Table~\ref{table:dynamics_parameters}. As per the original experimental paper \cite{Gregoire2013}, the CMOD was measured by the relative displacement in the $x$ direction of two points, $\SI{25}{\milli \metre}$ on either side of the midpoint of the beam. To apply the notch, all bonds that are at least partially within the volume of the void of the notch were broken as an initial condition. The force $P$ was measured as the vertical resultant force in the bonds of the displacement controlled particles.

\begin{table}[ht]
	\caption{Dynamics parameters for Example 2 and Example 3.}
	\centering
	\label{table:dynamics_parameters}
	\begin{tabular}{l c c}
		\toprule
	     & Example 2 & Example 3\\
		\midrule
		$\Delta t$ \SI{}{\second} & $1.135 \times 10^{-6}$ &$8.788 \times 10^{-7}$\\
    	$\eta$ \SI{}{\kilo\gram\per\metre\cubed\second} & $2.5 \times 10^{6}$ & $2.5 \times 10^{6}$\\
		max displacement rate \SI{}{\metre \per \second} & $3.5\times 10^{-10}$ & $8.0\times 10^{-8}$\\
		final displacement \SI{}{\metre} & $4.4\times 10^{-5}$& $2.7\times 10^{-3}$\\
		\bottomrule
	\end{tabular}
\end{table}

\smallskip
In this simple example, calibration of material parameters was performed on the unnotched experiment using a simple binary search. In order to find a point estimates of the bond stiffness $c$, the gradient of the elastic part of the experimental force-CMOD curve was calibrated to the gradient of the simulation force-displacement curve to two significant figures. In order to calibrate the trilinear parameters $s_0$, $s_1$ and $s_c$, a loss function in the form of a mean squared error between the experimental data and the numerical solution was minimised until the parameters converged to two significant figures through successive iterations of binary searches for each parameter.

\smallskip
The results for two possible calibrations, 'Point estimate UN 1' and 'Point estimate UN 2' (corresponding to two different minimum values of the loss function used) for the un-notched case are shown in Fig.~\ref{fig:calibration} (left). The values of the calibrated model parameters are compared to values from the trilinear model of \citet{Yang2018} in Table~\ref{table:optimal_parameters}. The calibrated model parameters were then tested on fifth-notched (FN)  beam geometry.

\begin{table}[ht]
	\caption{optimised parameters of the trilinear model against experimental data.}
	\centering
	\label{table:optimal_parameters}
	\begin{tabular}{l c c c}
		\toprule
	     &Trilinear model \citet{Yang2018} & Point Estimate 1 & Point Estimate 2\\
		\midrule
		$c$ & $2.32\times10^{18}$ & $1.792\times10^{18}$ & $1.792\times10^{18}$\\
		
    	$s_0$ &  $1.05\times 10^{-4}$ & $2.3\times 10^{-4}$ & $2.0\times 10^{-4}$\\
	
		$s_1$ & $6.90\times 10^{-4}$ & $2.42\times 10^{-4}$ & $2.42\times 10^{-4}$\\
		
		$s_c$ & $5.56\times10^{-3}$ & $1.45\times10^{-3}$ & $1.75\times10^{-3}$\\
		\bottomrule
	\end{tabular}
\end{table}

The trilinear model parameters determined by the calibration of the numerical simulation ('Point estimate UN 1' and 'Point estimate UN 2') are compared to the values determined by the trilinear model proposed by \citet{Yang2018} in Table~\ref{table:optimal_parameters}. The mean material properties of the experimental beam were used to calculate the trilinear model \cite{Yang2018} parameters: compressive strength $f_{cm,cyl} = $\SI{42.3}{\mega \pascal}; modulus of elasticity $E = $\SI{37.0}{\giga \pascal}; splitting tensile strength $f_t = $\SI{3.9}{\mega \pascal}. The density of the concrete mixture is $\rho = $\SI{2346}{\kilo \gram \per \metre \cubed}. The material fracture energy is determined empirically using \textit{fib} Model Code 2010: $G_F = 73f_{cm}^{0.18} = $\SI{143.2}{\newton \per \metre}.

\section{Example 3: Predicting the failure load of a reinforced concrete beam}\label{reinforced_concrete_beam}

This section presents the large-scale industry-motivated problem of estimating the failure load of a reinforced concrete beam. The numerical and experimental results for a reinforced concrete beam failing in shear, `Beam 5' from the series of reinforced concrete beam tests by \citet{Leonhardt1964}, commonly referred to as the Stuttgart Shear Tests.

\begin{figure}[H]
    \centering
    \includegraphics[width=14.0cm]{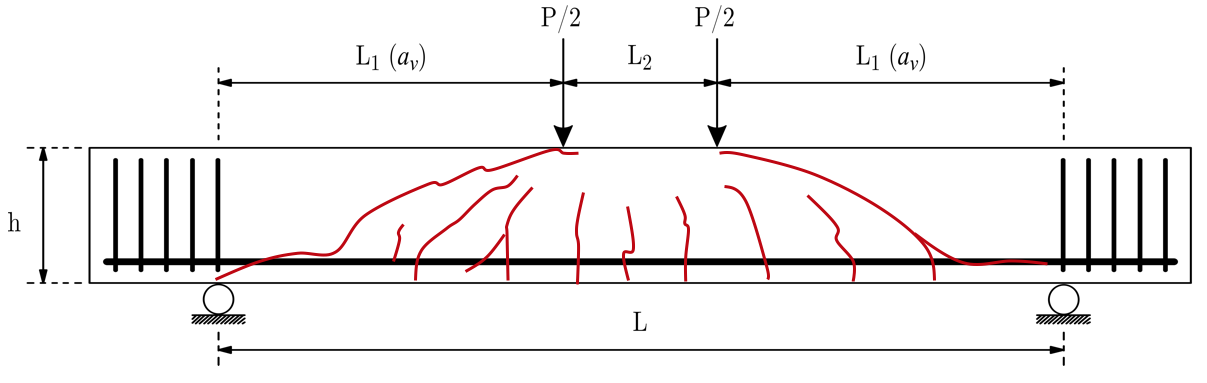}
    \caption{Schematic of the experimental setup. Adapted from \citet{Leonhardt1964}. Final fracture pattern of the experiment shown in red. }
    \label{fig:RC_experimental_setup}
\end{figure}

The beam is subjected to four-point loading, and shear reinforcement was not present between the applied loads so that the development of diagonal shear cracks could be studied. A schematic diagram of the experimental setup is illustrated in Fig.~\ref{fig:RC_experimental_setup}, overlaid by the fracture pattern observed towards the end of the test. The span of the member L = \SI{1950}{\milli \metre}, and the shear span \(a_v\) = \SI{810}{\milli \metre}. The effective depth d, the distance from the top compression fibre to the centre of the tensile reinforcement, is \SI{270}{\milli \metre}. The cross sectional area is \SI{320}{\milli \metre} \(\times\) \SI{190}{\milli \metre}, and the reinforcement ratio, the ratio between the area of steel \(A_s\) and the total cross-sectional area, is approximately 2.05\%. Longitudinal reinforcement is provided in the form of two ribbed steel bars with \SI{26}{\milli \metre} diameters. The reinforcing bars were manufactured from high-yield steel, for which we assume the following material properties: modulus of elasticity \(E_s\) = \SI{208000}{\mega \pascal}, yield strength \(f_y\) = \SI{465}{\mega \pascal}. Since the beam fails in shear, the steel was assumed not to yield, and hence it could be modelled with a linear damage model. Input parameters for the concrete constitutive model have been determined using empirical formulas from \citet{fib2010}, which relate the experimentally measured cubic compressive strength to common concrete properties. The following concrete properties were used: cubic compressive strength \(f_{cm,cube}\) = \SI{35}{\mega \pascal}, material fracture energy \(G_F\) = \SI{133}{\newton \per \metre} and modulus of elasticity \(E_c\) = \SI{30500}{\mega \pascal}.\smallskip

The peridynamic model was finely discretised using a $\Delta x = \SI{5}{\milli \metre}$ mesh with $\delta = \SI{15.7}{\milli \metre}$. The total number of peridynamic nodes was $n = 957,002$. A displacement-controlled loading scheme was used for the numerical simulations. Following the experimental setup closely, the beam was supported such that the right-hand support was constrained not to move in the vertical and out-of-plane directions. The left-hand support was clamped. The central supports were constrained not to move in the out-of-plane direction and had a prescribed displacement in the vertical direction. For the reasons outlined in Section~\ref{concrete_beams}, the boundary displacements were applied over two-by-two node cross-sections outside the surface of the beam at the positions of the arrows and supports on Fig.~\ref{fig:RC_experimental_setup}. These nodes were set as `no-failure-zones' so that they do not exceed critical stretch, which would be nonphysical, and their bond stiffness is the same as the bulk material. The displacement boundary conditions were applied smoothly over a fifth-order polynomial displacement time curve such that the boundary nodes had an acceleration of zero at the first time step and during crack nucleation and growth. This ensures that crack growth occurs under quasi-static loading conditions. The exact dynamic parameters used for the simulations and the maximum displacement rate of the applied displacement in the vertical direction are shown in Table~\ref{table:dynamics_parameters}.

\smallskip
In this simulation a bilinear damage model originally proposed by \citet{Gerstle2005} was used. \citet{Zaccariotto2015} determined the parameters of the bilinear damage model with material fracture energy $G_F$ and experimental load-displacement curves. The energy required to start the fracture process is defined as $G_0$ and $k_r = s_c/s_0 = G_f/G_0$. The stretch at the linear elastic limit $s_0$ given by
\begin{equation}\label{eq:bilinear_critical_stretch}
s_0 = \sqrt{\frac{5G_0}{6E\delta}}
\end{equation}
and $k_r$ is determined from experimental load-displacement curves or through sensitivity studies. Stiffness correction factors were applied to this numerical simulation. The force $P/2$ was measured as the vertical resultant force in the bonds of the displacement-controlled particles. The midspan deflection was measured as the average deflection of the nodes at the top of the beam at the midspan (\SI{975}{\milli \metre} from the left-hand support).

\begin{figure}
    \centering
    \includegraphics[width = 19.0cm]{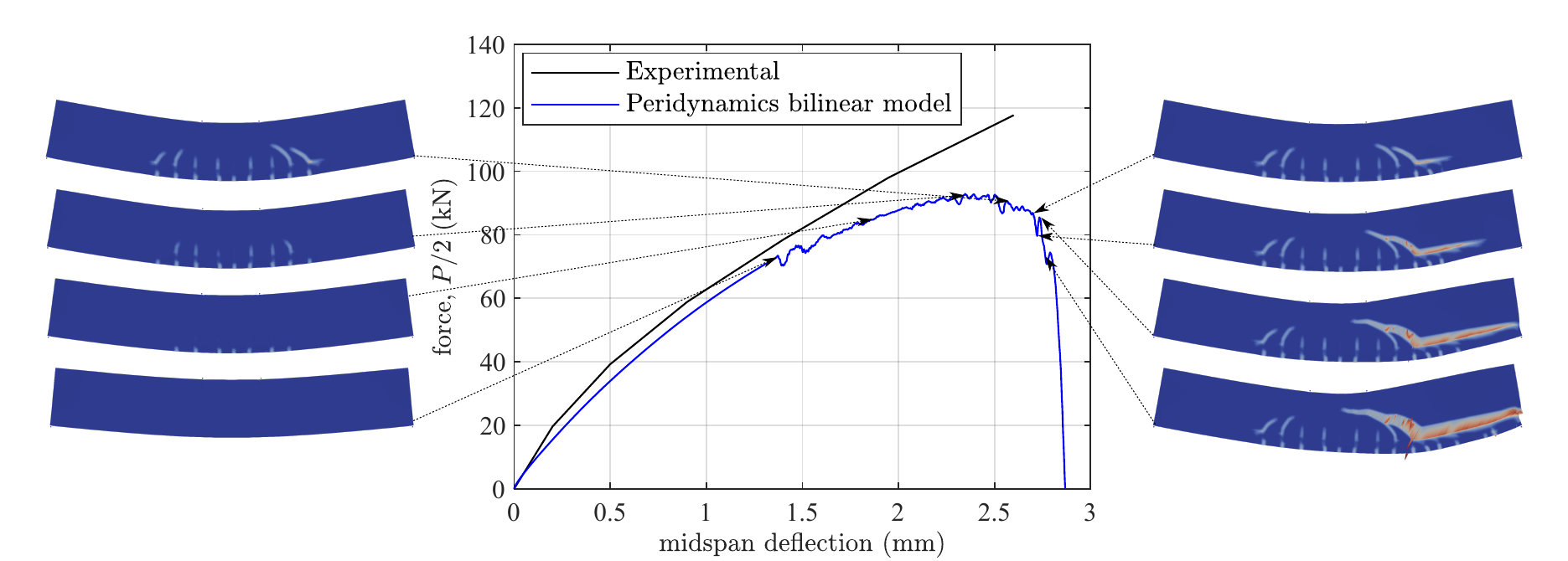}
    \caption{Comparison of load vs. midspan deflection for Beam 5 between experimental data and the bilinear peridynamic model results. The experimental data is from \citet{Leonhardt1964}. The deflection visualisation is amplified 50 times.}
    \label{fig:visual_force_displacement}
\end{figure}

\begin{figure}
    \centering
    \includegraphics[width = 0.8\linewidth]{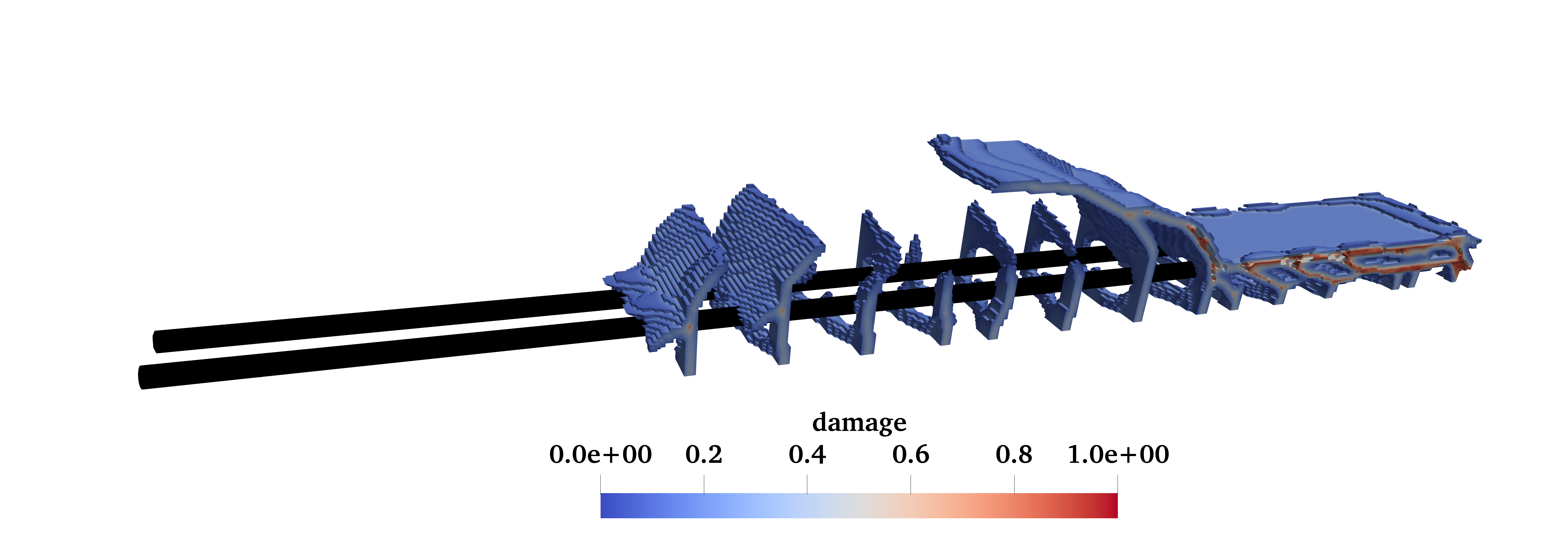}
    \caption{Threshold plot of peridynamic damage, showing three-dimensional visualisation of fracture pattern. The steel rebar nodes are contained in the black cylinders.}
    \label{fig:visual_cracks}
\end{figure}

A comparison between the force-deflection curve generated from model and experiment is shown in Fig.~\ref{fig:visual_force_displacement}, alongside inset plots show the evolution of the fracture pattern in the simulations. Without specific calibration of the damage model parameterisation for this test, the numerical and experimental model show good qualitative agreement. The numerical solution underestimates failure, and demonstrates a more progressive damage response to failure. The peridynamic model reproduces the expected asymmetrical failure mode, note that in both cases as the diagonal crack propagates into the compression zone, the characteristic rotation behaviour is seen. The final full crack is shown in a Paraview visualisation in Fig.~\ref{fig:visual_cracks}. Further improvements to the modelling of this specific experiment could be achieved by refining the model representation, application of loading and boundary conditions, and exploring different material models. Here however, the example demonstrates the scale of peridynamic solutions which can be readily solved using \texttt{PeriPy}.

\section{Future developments}\label{future}
This section outlines some possible extensions to the code. The user is encouraged to file bug reports or feature requests and make contributions by opening a `new issue' on GitHub (https://github.com/alan-turing-institute/PeriPy/issues).
\subsection{Stiffness correction algorithms}
The code has been developed to be easily extendable. Any partial volume algorithm, stiffness correction algorithm or micromodulus function can be applied with only minor changes to the code.\smallskip

\subsection{State-based Peridynamics}\label{sec:StateBased}
While \texttt{PeriPy} is currently a bond-based peridynamics solver, it may be extended using the same ideas to ordinary state-based (OSB) peridynamics. In OSB peridynamics, the bond force $\bm{f}_{ij}$ of one bond $\bm{\xi}_{ij}, j \in \mathcal{H}_i$ is not in general independent of another bond $\bm{\xi}_{ik}, j \in \mathcal{H}_i$ within the horizon. However, the individual components needed to calculate $\bm{f}_{ij}$ are calculated in two nested for-loops \cite{Littlewood2017}: one to calculate a `nodal dilation' using $\bm{\xi}_{ij}, j \in \mathcal{H}_i$ and $\bm{\eta}_{ij}$ and another to calculate the nodal force, also using $\bm{\xi}_{ij}, j \in \mathcal{H}_i$ and $\bm{\eta}_{ij}$. These two nested for-loops can be parallelised in the same way as the presented implementation, as the bond dilation contribution depends only on the state of the parent and child nodes. Since the bond dilation contributions cannot be factorised from the node force calculation, two kernels are now required to calculate the node force, which corresponds to the two aforementioned for-loops. Since this kernel contributes to 93-98\% of the overall compute time, we can expect the OSB implementation to be about two times slower than the presented bond-based implementation.

\subsection{Other}
Further contributions to this code include the extensions that have been discussed in literature, such as a contact algorithm for applying the boundary conditions; multiscale methods; and PD-FEM \cite{Kilic2010, Lee2017}, which is a natural extension to \texttt{PeriPy} given the mesh compatibility. Finally, it is believed that probabilistic methods for quantifying uncertainty in the peridynamic model parameters using real-life data will be a necessary for the application of the peridynamic method to industry problems and that \texttt{PeriPy} has opened up this avenue of research.

\section{Conclusion}\label{conclusions}
\texttt{PeriPy}, a lightweight, open-source and high-performance python package for solving peridynamics problems on a CPU/GPU platform in solid mechanics, is presented. The optimisation steps are discussed, namely, pairing a parallel computation of each bond force and a binary parallel reduction of the bond forces into the nodal force density by exploiting the execution and memory models of \texttt{OpenCL}, respectively. In all cases, \texttt{PeriPy} outperforms the existing CPU and GPU implementations. In particular, the optimisations lead to a 1.4-2.0 times speed increase compared to the existing \texttt{OpenCL} solver and a 3.7-7.3 times speed increase compared to the existing \texttt{OpenMP} solver. The implementation is compared on industry-motivated 3D benchmark problems with the other \texttt{OpenCL} and \texttt{OpenMP} solvers. The results indicate a step improvement in the algorithmic speed of peridynamic solvers. This work enables the use of parameter estimation, uncertainty quantification and outer-loop simulations in peridynamics research.
\appendix

\section*{Acknowledgements}
The authors would like to thank and acknowledge Jim Madge from the Alan Turing Institute for building the code base for this project and Greg Mingas from the Alan Turing Institute for his helpful comments and suggestions. The authors would also like to thank David Gr{\'{e}}goire for providing the plain concrete experimental data. This work was funded through TD's Turing AI Fellowship (2TAFFP/100007).

\section{Installation}\label{Installation}

Developers looking to contribute or make changes to the package should follow the installation instructions ``Get started from the GitHub repository (for developers)" on the GitHub page \href{https://github.com/alan-turing-institute/PeriPy}{\texttt{https://github.com/alan-turing-institute/PeriPy}}.

\smallskip
For users looking to create their own examples, the code is available to install at \href{https://pypi.org/project/peripy/}{\texttt{https://pypi.org/project/peripy/}}.

\smallskip
The latest documentation can be found at \href{https://peripy.readthedocs.io/en/latest/}{\texttt{https://peripy.readthedocs.io}}.

\section{Approximate maximum problem size}
\begin{table}[H]
	\caption{Approximate amount of global memory stored for a simulation with $n$ nodes and work-group size $N$.}
	\centering
	\label{table:memory_limitation}
	\begin{tabular}{l c}
		\toprule
		\text Features used in \texttt{peripy.model.Model} class & approximate total global memory used (in bytes)\\ 
		\midrule
		Single material and linear damage model, no stiffness corrections & $16n \times 8 + (N + 3)n \times 4$\\
		
    	Single material and linear damage model, stiffness corrections & $(16 + N)n \times 8 + (N + 3)n  \times 4$\\
	
		Composite material and/or n-linear damage model, no stiffness corrections & $16n \times 8 + (3N + 3)n   \times 4$\\
		
		Composite material and/or n-linear damage model, stiffness corrections & $(16 + N)n \times 8 + (3N + 3)n \times 4$\\
		\bottomrule
	\end{tabular}
\end{table}

\section{\texttt{OpenCL} Kernel Psuedocode}
\begin{lstlisting}[float=ht, style=CStyle, caption={\texttt{OpenCL} kernel for binary parallel reduction with sequential addressing}, label={lst:kernel}]
__kernel void
    reduction(
    __global double* body_force,
    __local double* local_cache_x,
    __local double* local_cache_y,
    __local double* local_cache_z,
    ){
        /* Calculate the force due to bonds on each node by binary parallel reduction.
         *
         * body_force - An (n,3) array of the current internal body forces of the particles
         * local_cache_x - local (local_size,) array to store the x components of the bond forces
         * local_cache_y - local (local_size,) array to store the y components of the bond forces
         * local_cache_z - local (local_size,) array to store the z components of the bond forces*/
        // Parallel reduction of the bond force onto node force
        // global_id is the bond number
        const int global_id = get_global_id(0);
        // local_id is the LOCAL node id in range [0, local_size) of a node in this parent node's family
        const int local_id = get_local_id(0);
        // local_size is N (usually 128 or 256, depending on the problem)
        const int local_size = get_local_size(0)
        // group_id is the node i
        const int node_id_i = get_group_id(0)
        for (int i = local_size/2; i > 0; i /= 2) {
            if(local_id < i) {
                local_cache_x[local_id] += local_cache_x[local_id + i];
                local_cache_y[local_id] += local_cache_y[local_id + i];
                local_cache_z[local_id] += local_cache_z[local_id + i];
            } 
            //Wait for all threads to catch up
            barrier(CLK_LOCAL_MEM_FENCE);
        }
        if (!local_id) {
            //Get the reduced forces
            body_force[3 * node_id_i + 0] = local_cache_x[0];
            body_force[3 * node_id_i + 1] = local_cache_y[0];
            body_force[3 * node_id_i + 2] = local_cache_z[0];
        }
    }
\end{lstlisting}

\begin{algorithm}[H]
\caption{Pseudocode of a parallel binary reduction across bond forces in an \texttt{OpenCL} kernel.}
\label{alg:reduction}
\begin{algorithmic}[1]
\ParFor{$i < n$}
\ParFor{$k < N$}
\Procedure{reduce\_bond\_forces}{$\bm{f}_{ij}$}
\State $j = \texttt{nlist}[iN + k]$ \Comment{Get neighbour}
\State $\texttt{local\_cache\_x}[k] = f_{x,ij}$ \Comment{Cache bond force data into local memory}
\State $\texttt{local\_cache\_y}[k] = f_{y,ij}$ \Comment{for each Cartesian direction}
\State $\texttt{local\_cache\_z}[k] = f_{z,ij}$
\Barrier \Comment{Wait for all threads to catch up.}
\EndBarrier
\For{$p = N/2; p > 0; p /= 2$}
\If{$k < p$}
\State $\texttt{local\_cache\_x}[k] += \texttt{local\_cache\_x}[k + p]$
\State $\texttt{local\_cache\_y}[k] += \texttt{local\_cache\_y}[k + p]$
\State $\texttt{local\_cache\_z}[k] += \texttt{local\_cache\_z}[k + p]$
\Barrier \Comment{Wait for all threads to catch up.}
\EndBarrier
\EndIf
\EndFor
\If{$k = 0$}
\State $F_{x,i} = \texttt{local\_cache\_x}[0]$
\State $F_{y,i} = \texttt{local\_cache\_y}[0]$
\State $F_{z,i} = \texttt{local\_cache\_z}[0]$
\EndIf
\EndProcedure
\EndParFor
\EndParFor
\end{algorithmic}
\end{algorithm}

\begin{algorithm}[H]
\caption{Calculation of bond forces in an \texttt{OpenCL} kernel parallelising over bonds and nodes.}
\label{alg:bond_index}
\begin{algorithmic}[1]
\ParFor{$i < n$}
\ParFor{$k < N$}
\State $j = \texttt{nlist}[iN + k]$
\If{$j \neq -1$} \Comment{Bond is not broken}
\State $\bm{\xi}_{ij} = \bm{x}_j - \bm{x}_i$
\State $\bm{\eta}_{ij} = \bm{u}_j - \bm{u}_i$
\State $s = \frac{\|\bm{\eta} + \bm{\xi}\| - \| \bm{\xi}\|}{\| \bm{\xi}\|}$
\If{$s < s_0$}
\State $\bm{f}_{ij} = \frac{\bm{\eta} + \bm{\xi}}{\|\bm{\eta} + \bm{\xi}\|} s \lambda_{ij} c \beta_{ij} V_j \mu_{ij}$
\Else \Comment{Break bond}
\State $\texttt{nlist}[iN + k] = -1$
\State $\bm{f}_{ij} = \bm{0}$
\EndIf
\Else \Comment{Bond is broken}
\State $\bm{f}_{ij} = \bm{0}$
\EndIf
\Procedure{reduce\_bond\_forces}{$\bm{f}_{ij}$}
\EndProcedure
\EndParFor
\EndParFor
\end{algorithmic}
\end{algorithm}

\begin{algorithm}[H]
\caption{Calculation of bond forces in an \texttt{OpenCL} kernel by parallelising over nodes.}
\label{alg:node_index}
\begin{algorithmic}[1]
\ParFor{$i < n$}
\State $F_{x, i} = 0$\algorithmiccomment{Initiate nodal forces}
\State $F_{y, i} = 0$
\State $F_{z, i} = 0$
\For{$k < N$}
\State $j = \texttt{nlist}[iN + k]$
\If{$j \neq -1$} \algorithmiccomment{Bond is not broken}
\State $\bm{\xi}_{ij} = \bm{x}_j - \bm{x}_i$
\State $\bm{\eta}_{ij} = \bm{u}_j - \bm{u}_i$
\State $s = \frac{\|\bm{\eta} + \bm{\xi}\| - \| \bm{\xi}\|}{\| \bm{\xi}\|}$
\State $\bm{f}_{ij} = \frac{\bm{\eta} + \bm{\xi}}{\|\bm{\eta} + \bm{\xi}\|} s \lambda_{ij} c \beta_{ij} V_j \mu_{ij}$
\State $F_{x, i} += f_{x, ij}$\algorithmiccomment{Reduce bond forces with a serial summation}
\State $F_{y, i} += f_{y, ij}$
\State $F_{z, i} += f_{z, ij}$
\EndIf
\EndFor
\EndParFor
\end{algorithmic}
\end{algorithm}

\begin{algorithm}[H]
	\caption{Pseudocode of velocity-Verlet scheme in an \texttt{OpenCL} kernel.}
	\label{alg:velocity_verlet}
	\begin{algorithmic}[1]
		\Procedure{Update}{displacements and velocities}
		\ParFor{$i < n$}
		\State $\bm{\dot{u}}_{i}(t + \frac{\Delta t}{2}) = \bm{\dot{u}}_{i}(t) + \frac{\Delta t}{2} \bm{\ddot{u}}_{i}(t)$ \Comment{Half-step velocity.}
		\State $\bm{\ddot{u}}_{i}(t + \Delta t) = \frac{\bm{f}_{i}(t)  - \eta \bm{\dot{u}}_{i}(t + \frac{\Delta t}{2})}{\rho_{i}}$
		\State $\bm{u}_{i}(t + \Delta t) = \bm{u}_{i}(t) + \Delta t \bm{\dot{u}}_{i}(t) + \frac{\Delta t^{2}}{2} \bm{\ddot{u}}_{i}(t)$
        \State $\bm{\dot{u}}_{i}(t+ \Delta t) = \bm{\dot{u}}_{i}(t) + \frac{\Delta t}{2}\left(\bm{\ddot{u}}_{i}(t) + \bm{\ddot{u}}_{i}(t + \Delta t)\right)$
        \State $t = t + \Delta t$ \Comment{time-step update.}
        \EndParFor
		\EndProcedure
	\end{algorithmic}
\end{algorithm}

\bibliography{main}

\end{document}